\documentclass{emulateapj}
\usepackage{natbib}
\usepackage{apjfonts}
\begin{document}

\title{The Black Hole in the Compact, High-dispersion Galaxy NGC 1271}

\author{Jonelle L. Walsh$^{1,2}$, Remco C.~E. van den Bosch$^{3}$,
  Karl Gebhardt$^{2}$, Ak{\i}n Y{\i}ld{\i}r{\i}m$^{3}$, Kayhan
  G\"{u}ltekin$^{4}$, Bernd Husemann$^{5,6}$, {\sc and} Douglas
  O. Richstone$^{4}$}

\affil{$^1$ George P. and Cynthia Woods Mitchell Institute for
  Fundamental Physics and Astronomy, and Department of Physics and
  Astronomy, Texas A\&M University, College Station, TX 77843, USA;
  walsh@physics.tamu.edu \\
  $^2$ Department of Astronomy, The University of Texas at Austin,
  2515 Speedway, Stop C1400, Austin, TX 78712, USA \\
  $^3$ Max-Planck Institut f\"{u}r Astronomie, K\"{o}nigstuhl 17,
  D-69117 Heidelberg, Germany\\
  $^4$Department of Astronomy, University of Michigan, 1085
  S.~University Ave., Ann Arbor, MI 48109, USA\\
  $^5$European Southern Observatory, Karl-Schwarzschild-Str. 2, 85748
  Garching, Germany\\
  $^{6}$Leibniz Institute for Astrophysics Potsdam, An der Sternwarte
  16, 14482 Potsdam, Germany}

\begin{abstract}

  Located in the Perseus cluster, NGC 1271 is an early-type galaxy
  with a small effective radius of 2.2 kpc and a large bulge stellar
  velocity dispersion of 276 km s$^{-1}$ for its $K$-band luminosity
  of $8.9\times10^{10}\ L_\odot$. We present a mass measurement for
  the black hole in this compact, high-dispersion galaxy using
  observations from the integral field spectrograph NIFS on the Gemini
  North telescope assisted by laser guide star adaptive optics,
  large-scale integral field unit observations with PPAK at the Calar
  Alto Observatory, and \emph{Hubble Space Telescope} WFC3 imaging
  observations. We are able to map out the stellar kinematics both on
  small spatial scales, within the black hole sphere of influence, and
  on large scales that extend out to four times the galaxy's effective
  radius. We find that the galaxy is rapidly rotating and exhibits a
  sharp rise in the velocity dispersion. Through the use of
  orbit-based stellar dynamical models, we determine that the black
  hole has a mass of $(3.0^{+1.0}_{-1.1}) \times 10^9\ M_\odot$ and
  the $H$-band stellar mass-to-light ratio is $1.40^{+0.13}_{-0.11}\
  \Upsilon_\odot$ ($1\sigma$ uncertainties). NGC 1271 occupies the
  sparsely-populated upper end of the black hole mass distribution,
  but is very different from the Brightest Cluster Galaxies (BCGs) and
  giant elliptical galaxies that are expected to host the most massive
  black holes. Interestingly, the black hole mass is an order of
  magnitude larger than expectations based on the galaxy's bulge
  luminosity, but is consistent with the mass predicted using the
  galaxy's bulge stellar velocity dispersion. More compact,
  high-dispersion galaxies need to be studied using high spatial
  resolution observations to securely determine black hole masses, as
  there could be systematic differences in the black hole scaling
  relations between these types of galaxies and the BCGs/giant
  ellipticals, thereby implying different pathways for black hole and
  galaxy growth.

\end{abstract}


\section{Introduction}
\label{sec:intro}

Over the past 15 years, it has become increasingly clear that
supermassive black holes are an essential component of a galaxy, as
demonstrated by the correlations connecting black hole masses and
galaxy bulge properties (e.g., \citealt{Kormendy_Richstone_1995,
  Ferrarese_2000, Gebhardt_2000, Marconi_Hunt_2003, Gultekin_2009,
  Kormendy_Ho_2013}). Supermassive black holes are thought to regulate
galaxy properties and influence star formation via feedback processes
\citep{Silk_Rees_1998, Fabian_1999}, however the black hole -- bulge
relations can also arise because of the inherent averaging associated
with random galaxy mergers, without the need for the black hole to
actively influence its host galaxy \citep{Peng_2007,
  Jahnke_Maccio_2011}. Roughly 80 dynamical black hole mass
($M_\mathrm{BH}$) measurements have been made to date
\citep{Kormendy_Ho_2013}, almost exclusively through the use of high
angular resolution facilities such as the \emph{Hubble Space
  Telescope} (\emph{HST}) and 8$-$10m ground-based telescopes with the
aid of adaptive optics (AO). Despite the growing number of black hole
mass measurements, the local black hole mass census is highly
incomplete. Gaining a more complete picture of black hole demographics
and a deeper understanding the mechanisms that drive black hole/galaxy
evolution requires the secure measurement of many more black holes,
particularly those at the extremes of the black hole mass scale and in
a wider range of galaxy types.

With the goal of finding more objects suitable for future dynamical
black hole mass measurements, the HET Massive Galaxy Survey obtained
long-slit spectra of $\sim$1000 galaxies using the Hobby-Eberly
Telescope (HET) at McDonald Observatory \citep{vandenBosch_2015}. The
survey uncovered a sample of interesting early-type galaxies with
small sizes and high stellar velocity dispersions for their
luminosities. More quantitatively, these galaxies have an effective
radius, $R_e$, below 3 kpc, a central stellar velocity dispersion
larger than $250$ km s$^{-1}$, and $K$-band luminosities
$\sim$$(5-25)\times10^{10}\ L_\odot$. The HET spectra hint that these
compact, high-dispersion galaxies could host some of the largest black
holes known ($M_\mathrm{BH} > 10^9\ M_\odot$), and that the black
holes could weigh a high fraction of its host galaxy's mass. Six
example objects were highlighted in \cite{vandenBosch_2012}, and
orbit-based stellar dynamical models were calculated for one galaxy,
NGC 1277. In the case of NGC 1277, \cite{vandenBosch_2012} found a
$1.7\times10^{10}\ M_\odot$ black hole that is surprisingly 59\% of
the galaxy's bulge mass, or 14\% of the galaxy's total mass. While
\cite{Yildirim_2015} infer a similar $M_\mathrm{BH}$ from
seeing-limited, large-scale integral field unit (IFU) data,
\cite{Emsellem_2013} show a smaller black hole of a few billion solar
masses can also reasonably reproduce the observed HET long-slit
kinematics presented in \cite{vandenBosch_2012}.

Obtaining secure black hole mass measurements for the other compact,
high-dispersion galaxies found through the HET Massive Galaxy Survey
is important for addressing questions concerning the upper end of the
black hole -- host galaxy relationships. With the present sample of
black hole mass measurements, the slope, intrinsic scatter, and even
the shape of the correlations for high-mass black holes are not well
established (e.g., \citealt{McConnell_Ma_2013}). Also, the black hole
mass -- stellar velocity dispersion relation ($M_\mathrm{BH}$ --
$\sigma_\star$) and the black hole mass -- bulge luminosity relation
($M_\mathrm{BH}$ -- $L_\mathrm{bul}$) are in direct conflict at the
upper end, and make drastically different predictions for the inferred
number density of the most massive black holes \citep{Lauer_2007}. Not
only are the compact, high-dispersion galaxies from the HET survey
useful for filling in the poorly sampled high-mass end of the black
hole relations, but also the mass estimates from $M_\mathrm{BH}$ --
$\sigma_\star$ and $M_\mathrm{BH}$ -- $L_\mathrm{bul}$ differ by a
factor of at least three. Consequently, the galaxies are also useful
for testing which of the correlations is more fundamental and a better
predictor of $M_\mathrm{BH}$ at the high-mass end of the scaling
relations.

While recent progress has been made in searching for, and revising
measurements for, black holes with masses larger than $10^9\ M_\odot$
\citep{Shen_2010, Gebhardt_2011, McConnell_2011, McConnell_2012,
  vandenBosch_2012, Walsh_2010, Walsh_2013, Rusli_2013}, many of these
galaxies are giant ellipticals or Brightest Cluster Galaxies (BCGs),
which are often large (with $R_e > 10$ kpc; e.g.,
\citealt{DallaBonta_2009}), have cored surface brightness profiles,
and are dispersion-supported showing little to no rotation. In
contrast, the compact, high-dispersion galaxies found through the HET
survey are small, rapidly rotating, and generally exhibit cuspy
surface brightness profiles. Such host galaxy environments haven't
been extensively explored on the black hole -- host galaxy
relationships. Besides NGC 1277, only the compact galaxies NGC 1332
\citep{Rusli_2011}, NGC 4342 \citep{Cretton_vandenBosch_1999}, NGC
4486B \citep{Kormendy_1997}, and M60-UCD1 \citep{Seth_2014} have
dynamical black hole mass measurements, with NGC 1332 and NGC 4342
being most like NGC 1277. For these galaxies, the black hole mass
measurements are in agreement with $M_\mathrm{BH}$ -- $\sigma_\star$
given the intrinsic scatter of the relation, but are positive outliers
from the $M_\mathrm{BH}$ -- $L_\mathrm{bul}$ relation
\citep{Kormendy_Ho_2013}. We note that there are uncertainties
associated with the bulge luminosity for NGC 1332 (see
\citealt{Kormendy_Ho_2013} for details) and the black hole mass
measurement for NGC 4486B (see \citealt{Gultekin_2009} for
details). Also, tidal stripping is believed to be the cause of the
over-massive black hole in the ultracompact dwarf galaxy M60-UCD1
\citep{Seth_2014}, and there is some debate as to whether NGC 4486B
and NGC 4342 have been stripped as well (e.g., \citealt{Faber_1973,
  Bogdan_2012, Blom_2014}). Nevertheless, additional similar galaxies
need to be studied because there could be systematic differences in
the scaling relations between the compact, high-dispersion galaxies
and the giant ellipticals/BCGs. If so, that would imply that the black
holes in the two types of galaxies grew in different ways.

While the compact, high-dispersion galaxies are unusual in the
present-day Universe, they are qualitatively similar to the typical
$z\sim2$ quiescent galaxies, which are also small, have disk-like
features, and could have high velocity dispersions \citep{Zirm_2007,
  vanDokkum_2009, vanderWel_2011}. The $z\sim2$ red nuggets are
believed to be the progenitors of the massive early-type galaxies seen
today, evolved in size and mass (e.g.,
\citealt{vanDokkum_2010}). Thus, the compact, high-dispersion galaxies
found through the HET survey could provide clues to the link between
local galaxies and the $z\sim2$ red nuggets, the ultra compact sub-mm
galaxies at $z\sim3$ \citep{Toft_2014}, and the early massive black
holes found in $z>6$ quasars \citep{Fabian_2013}.

We have begun to obtain the imaging and spectroscopic observations
necessary for a detailed examination of the compact, high-dispersion
galaxies from the HET Massive Galaxy Survey. This includes \emph{HST}
and AO-assisted IFU observations to probe the region over which the
black hole dominates the galaxy's potential (the black hole sphere of
influence; $r_\mathrm{sphere} = G M_\mathrm{BH}/\sigma_\star^2$), and
IFU observations that sample the large-scale stellar kinematics out to
several effective radii. In this paper, we focus on measuring the mass
of the black hole for the first compact, high-dispersion galaxy for
which we have completed AO IFU observations. NGC 1271 has not been
widely investigated in the literature and is given an uncertain SB0
classification according to the NASA/IPAC Extragalactic Database
(NED). The galaxy is located within the Perseus cluster, at
$z=0.0192$, and we adopt a distance to NGC 1271 of 80 Mpc, which is
the Hubble flow distance derived from the \cite{Mould_2000} Virgo +
Great Attractor + Shapley Supercluster Infall velocity field model
assuming a Hubble constant of $H_0 = 70.5$ km s$^{-1}$ Mpc$^{-1}$, a
matter density of $\Omega_M = 0.27$ and a cosmological constant of
$\Omega_\Lambda = 0.73$. The Sloan Digital Sky Survey $g-i$ color is
1.6 and absolute $r$-band magnitude is -20.8 for the galaxy. Long-slit
spectra of NGC 1271 were obtained through the HET Massive Galaxy
Survey, and the reported [\ion{N}{2}]/H$\alpha$ and
[\ion{O}{3}]/H$\beta$ emission-line ratios measured within a 3\farcs5
aperture \citep{vandenBosch_2015} place NGC 1271 just within the
composite galaxies section near the active galactic nuclei side of
\cite{Kewley_2006}. Also, \cite{vandenBosch_2015} do not find the
presence of any broad emission lines.

In Section \ref{sec:obs}, we describe the imaging and spectroscopic
observations, including the data reduction procedures. We present the
luminous mass model for the galaxy in Section
\ref{sec:stellarmassmodel} and the stellar kinematics in Section
\ref{sec:stellarkin}. Our determination of the point-spread function
(PSF) for the spectroscopic observations is discussed in Section
\ref{sec:psf}, and an overview of the orbit-based stellar dynamical
models is given in Section \ref{sec:models}. In Section
\ref{sec:results}, we present the results of the models and examine
possible sources of systematic uncertainty. Finally, in Sections
\ref{sec:discussion} and \ref{sec:conclusion}, we study the galaxy's
orbital structure, discuss NGC 1271 within the context of the
$M_\mathrm{BH}$ -- host galaxy relationships, and summarize our
findings.

\section{Observations}
\label{sec:obs} 

For NGC 1271, we obtained imaging observations with the \emph{HST}
Wide-Field Camera 3 (WFC3) in order to measure the galaxy's surface
brightness distribution. We also acquired spectra with the
Near-infrared Integral Field Spectrometer (NIFS;
\citealt{McGregor_2003}) on the 8.1m Gemini North telescope assisted
by the ALTtitude conjugate Adaptive optics for the InfraRed
\citep{Herriot_2000, Boccas_2006} system. The NIFS data is important
for constraining $M_\mathrm{BH}$, as it resolves the black hole sphere
of influence. We also acquired large-scale spectra with the Postdam
Multi Aperture Spectrograph (PMAS; \citealt{Roth_2005}) in the Pmas
fiber PAcK (PPAK; \citealt{Verheijen_2004, Kelz_2006}) mode at the
3.5m telescope at Calar Alto Observatory. Although long-slit
spectroscopic observations along the galaxy major axis have been
previously made using the HET, measuring a large-scale,
two-dimensional (2D) velocity field is preferable over a single slit
observation for constraining the stellar mass-to-light ratio and the
stellar orbital distribution. Hence, we use the PPAK IFU observations
in place of the major-axis HET measurements. Below we describe the
WFC3, NIFS, and PPAK observations and data reduction methods.

\subsection{\emph{HST} Imaging}
\label{subsec:hstimag}

We observed NGC 1271 with \emph{HST} WFC3 and the IR/F160W filter
under program GO-13050. The observation was composed of three dithered
full array exposures of 450 s, and four dithered subarray exposures of
1.7 s, leading to a total integration time of 1354 s. The short
subarray exposures were chosen to ensure the nucleus would not become
saturated. The flattened, calibrated images were corrected for
geometric distortions, cleaned, and combined using AstroDrizzle
\citep{Gonzaga_2012}. Since the exposures were dominated by galaxy
light, we found that the standard AstroDrizzle sky subtraction
overestimated the background flux. We therefore manually measured the
background level in each of the images. For the full array images, we
measured the flux from the corners of each image, while for the
subarray exposures we measured the flux difference between the
sky-subtracted full frames and the subarray frames. With the
background level determined, the exposures were combined to produce a
super-sampled image with a spatial resolution of 0\farcs06
pixel$^{-1}$. Not only is the \emph{HST} image suitable for
determining the luminous distribution on near the black hole, but due
to the small size of NGC 1271, we were also able to measure the
luminosity out to larger galaxy scales, extending to $\sim$5 $R_e$
(adopting an effective radius of 5\farcs6, or 2.2 kpc, measured from a
single component S\'{e}rsic fit to the \emph{HST} F160W image; see
Section \ref{subsec:bhrels}).

\subsection{NIFS Spectroscopy}
\label{subsec:obsnifs}

The NIFS laser guide star (LGS) AO observations were acquired over
three nights, on 2012 Dec 27, 2012 Dec 29, and 2013 Jan 8, in queue
mode under program GN-2012B-Q-51. We used the $H+K$ filter and the $K$
grating with a central wavelength of 2.2 $\mu$m to obtain spectra over
a 3\arcsec$\times$3\arcsec\ field-of-view and a spectral resolution of
R$\sim$5290. We recorded 900 s exposures of the galaxy nucleus,
following an Object-Sky-Object observing sequence, totaling 3 hours of
on-source integration. An $R=17.8$ mag star located 17\arcsec\ away
from the galaxy was used as the tip-tilt reference. In addition, we
observed the tip-tilt star to monitor the PSF during each of the three
nights and the A0~V stars HIP 10559 and HIP 22842 for telluric
correction.

The data were reduced using IRAF\footnote[7]{IRAF is distributed by
  the National Optical Astronomy Observatory, which is operated by the
  Association of Universities for Research in Astronomy under
  cooperative agreement with the National Science Foundation} tasks
within the Gemini/NIFS package version 1.11, utilizing the example
NIFS processing
scripts\footnote[8]{http://www.gemini.edu/sciops/instruments/nifs/data-format-and-reduction}. The
reduction included sky subtraction, flat fielding, interpolation over
bad pixels, cosmic-ray cleaning, and spatial rectification and
wavelength calibration using Ronchi mask and arc lamp exposures. The
spectra were then corrected for telluric features, using an A0~V star
spectrum, after interpolating over the Br$\gamma$ absorption line and
dividing by a black body with a temperature of 9480 K. Next, a data
cube was produced, having $x$ and $y$ spatial dimensions, each with a
scale of 0\farcs05 pixel$^{-1}$, and one spectral dimension,
$\lambda$, using a common wavelength range and sampling for the
individual science exposures. The relative spatial positions between
the data cubes were determined by summing along the wavelength axis
and cross-correlating the resulting flux maps. The offsets were used
to align and combine the 12 individual exposures, generating the final
data cube of the galaxy. Data reduction of the PSF star observations
followed a similar procedure.

\subsection{PPAK Spectroscopy}
\label{subsec:obsppsk}

The PPAK observations of NGC 1271 were acquired as part of a campaign
to obtain large-scale spectroscopy of the compact, high-dispersion
galaxies. The observations, data reduction, and kinematic measurements
for NGC 1271 will be presented in \cite{Yildirim_inprep}, but follow
closely the PPAK observations for two other compact, high-dispersion
galaxies described in \cite{Yildirim_2015}. For completeness, we
briefly review the pertinent information below and in Section
\ref{subsec:ppakkin}.

The wide-field IFU observations were taken over three nights, from
2013 Jan 4-6, using the V500 grating to provide coverage of
$4200-7000$ \AA\ with a spectral resolution of R$\sim$850 at 5000
\AA. We used three dithers to fully sample the 331 2\farcs7-diameter
fibers, to increase the spatial resolution of the data, and to address
effects of vignetting and bad pixels. An additional 36 fibers, located
72\arcsec\ away from the center of the instrument field-of-view, were
used to measure the sky. During each of the three nights, two 1200 s
science exposures were taken at each of the three dither positions,
leading to a total of 6 hours of on-source integration. The data
reduction followed the procedure adopted for the Calar Alto Legacy
Integral Field Spectroscopy Area Survey. The main steps included bias
subtraction, flat-fielding, cosmic ray cleaning, extraction of
spectra, wavelength calibration, sky subtraction, and flux calibration
using spectrophotometric standard stars. Spectra from the three
pointings were then combined and resampled into a data cube, followed
by a correction for differential atmospheric refraction. We note that
the line spread function is measured as a function of wavelength and
fiber position from arc lamps during the wavelength calibration step,
and then homogenized to a common value prior to the extraction of the
PPAK kinematics. The details of these steps are discussed at length by
\cite{Sanchez_2012} and \cite{Husemann_2013} and we refer the reader
to those publications for additional information.

\section{Constructing the Luminous Mass Model}
\label{sec:stellarmassmodel}

We generated a luminous mass model for NGC 1271 by parameterizing the
\emph{HST} WFC3 F160W image as the sum of 2D Gaussians using the
Multi-Gaussian Expansion (MGE) formalism \citep{Monnet_1992,
  Emsellem_1994}. The MGE method is able to reproduce a wide range of
galaxy surface brightness profiles and allows for an analytic
deprojection to determine the intrinsic luminosity density. Here, we
use the image decomposition package Galfit \citep{Peng_2010} because
it takes into consideration an error map during the fit and allows for
the detailed examination of model residuals, but we utilize the
implementation of \cite{Cappellari_2002} to determine suitable
starting parameter values for the initial run with Galfit. When
constructing the MGE model, we account for the WFC3 PSF, which we
adopt from \cite{vanderWel_2012}. This PSF was generated with Tiny Tim
\citep{Krist_Hook_2004} for the F160W filter at the center of the WFC3
detector assuming a G2 V spectral type, and drizzled to produce a PSF
with the same spatial scale as our final science image of NGC 1271. In
addition, we identified foreground objects using the program
SExtractor \citep{Bertin_Arnouts_1996} and masked these regions during
the MGE fit.

\begin{figure}
\begin{center}
\epsscale{1.0}
\plotone{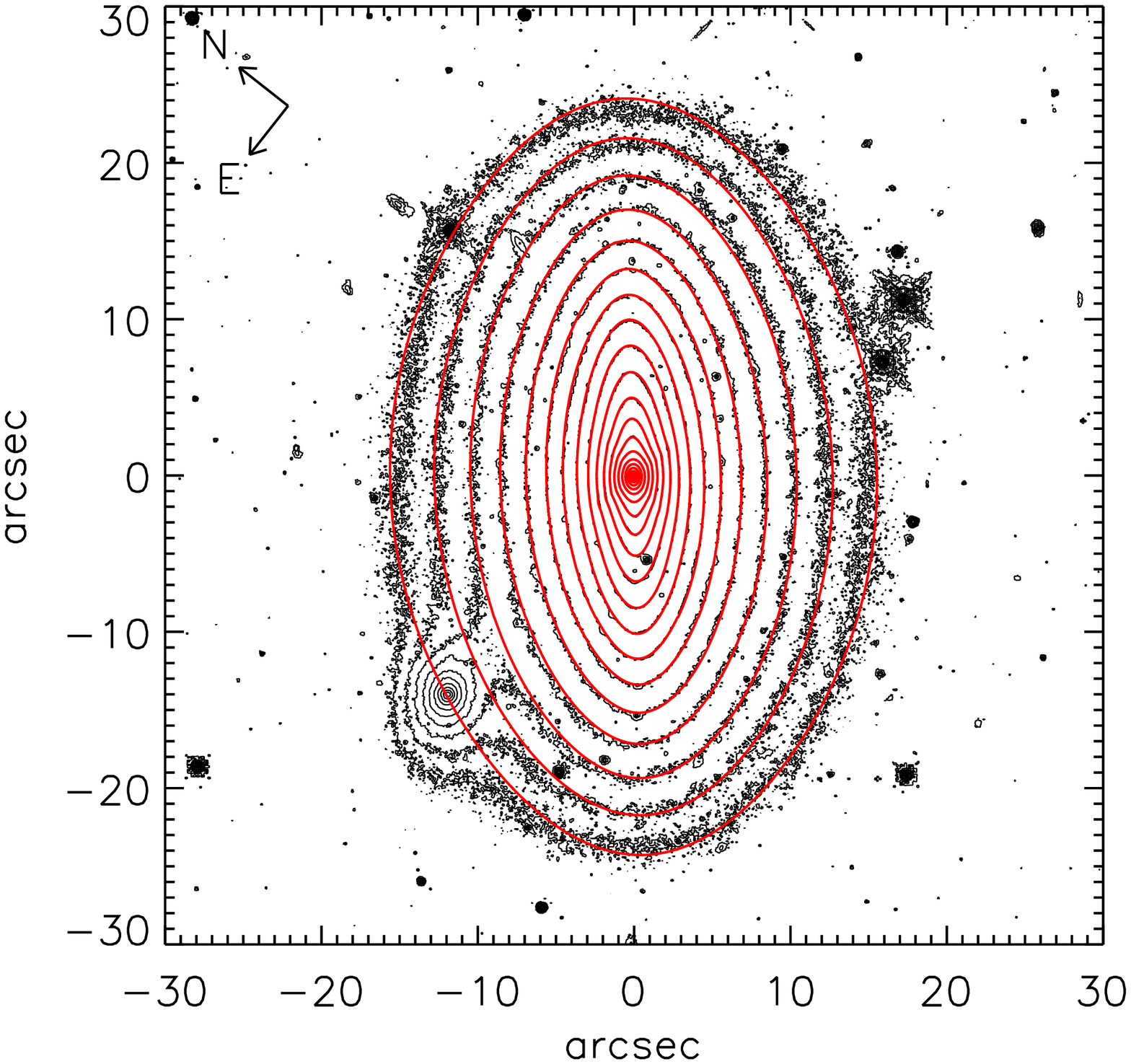} \\
\plotone{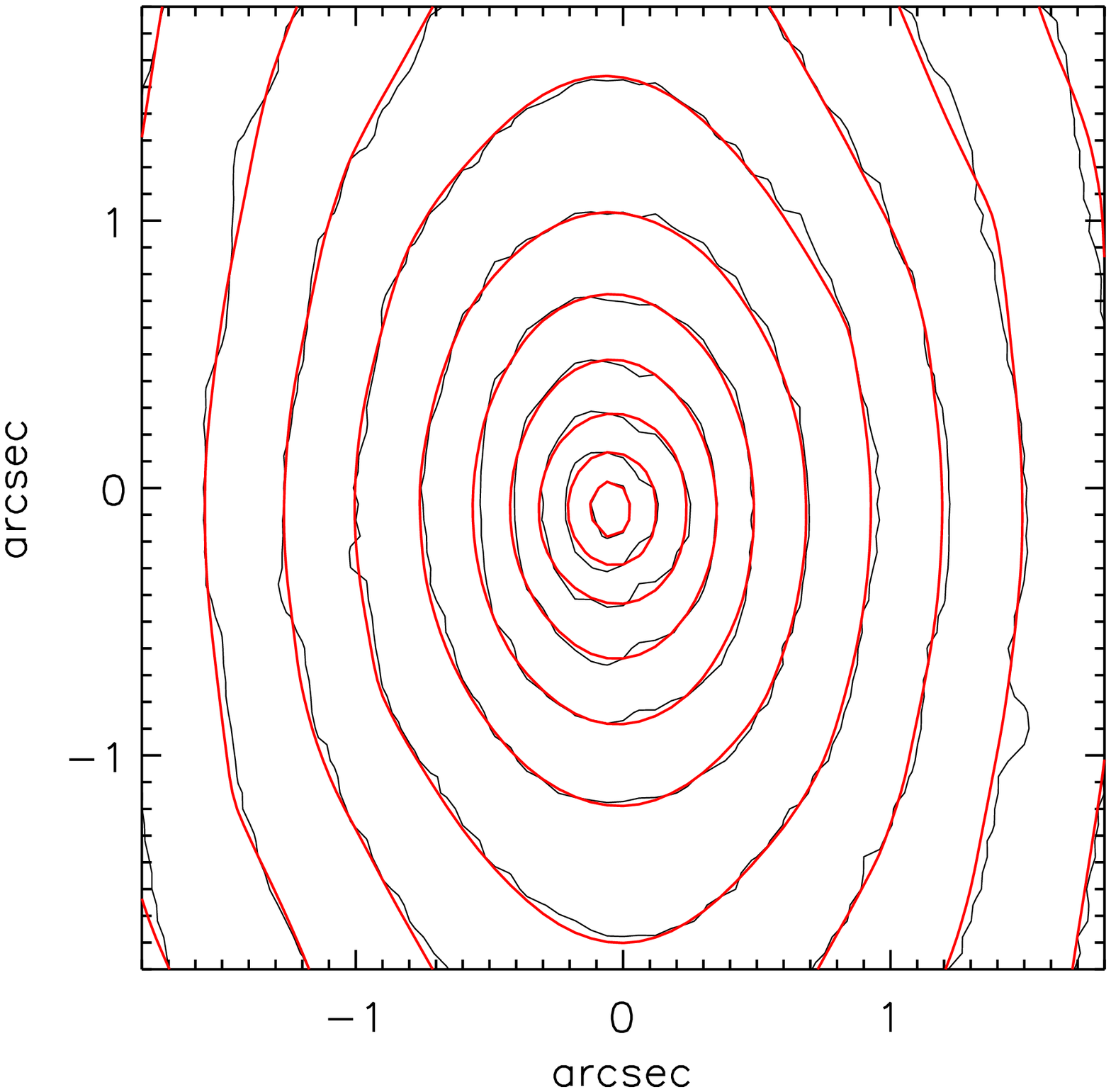}
\caption{Isophotes of the MGE model (red) are compared to the
  \emph{HST} WFC3 F160W image (top) and to the inner
  3\farcs6$\times$3\farcs6 region of the image (bottom). Contours are
  logarithmically spaced, but arbitrary. Foreground stars and galaxies
  were masked during the MGE fit. \label{fig:mge}}
\end{center}
\end{figure}

The final MGE model contained 11 components, where each Gaussian was
set to have the same position angle and center and the projected axis
ratio ($q^\prime$) was required to be larger than 0.25. Due to the
degeneracy associated with fitting a large number of Gaussians, we
chose to restrict $q^\prime > 0.25$ in order to avoid highly flattened
components that place very stringent constraints on the viewing angles
for which a model can be deprojected. The final MGE model is an good
representation of the galaxy, as can be seen in Figure
\ref{fig:mge}. We present the final MGE parameters, after correction
for galactic extinction using the \cite{Schlafly_Finkbeiner_2011}
value given by NED, in Table \ref{tab:mge}.

\begin{deluxetable}{lcccccccc}
\tabletypesize{\scriptsize}
\tablewidth{0pt}
\tablecaption{MGE Parameters \label{tab:mge}}
\tablehead{
\colhead{$j$} & 
\colhead{} &
\colhead{$\log\ I_j$ (L$_{\odot, H}$ pc$^{-2}$) } &
\colhead{} &
\colhead{$\sigma_j^\prime$ (\arcsec)} &
\colhead{} &
\colhead{} &
\colhead{} &
\colhead{$q_j^\prime$} \\
\colhead{(1)} &
\colhead{} &
\colhead{(2)} &
\colhead{} &
\colhead{(3)} &
\colhead{} &
\colhead{} &
\colhead{} &
\colhead{(4)}
}

\startdata

1   &&   6.074  &&   0.073  &&&&   0.25 \\
2   &&   4.778  &&   0.175  &&&&   0.83 \\
3   &&   4.786  &&   0.399  &&&&   0.56 \\
4   &&   4.183  &&   0.765  &&&&   0.77 \\
5   &&   3.172  &&   1.838  &&&&   0.66 \\
6   &&   3.756  &&   2.018  &&&&   0.26 \\
7   &&   3.391  &&   4.306  &&&&   0.25 \\
8   &&   3.054  &&   5.857  &&&&   0.38 \\
9   &&   2.474  &&   8.876  &&&&   0.49 \\
10  &&   1.879  &&  12.946  &&&&   0.72 \\
11  &&   0.889  &&  24.772  &&&&   0.99

\enddata

\tablecomments{The component number is listed in column (1), the
  central surface brightness, using a galactic extinction of 0.085 and
  a solar absolute magnitude of 3.33, is provided in column (2), the
  dispersion along the major axis is given in column (3), and the axis
  ratio is presented in column (4). The components all have of a
  position angle of $-50.3^\circ$ and projected quantities are denoted
  with primed variables.}

\end{deluxetable}

\section{Extracting the Stellar Kinematics}
\label{sec:stellarkin}

From the NIFS and PPAK data cubes, we measured the line-of-sight
velocity distribution (LOSVD) as a function of spatial location. The
LOSVD was described using the first four Gauss-Hermite (GH) moments:
the radial velocity ($V$), the velocity dispersion ($\sigma$), and
$h_3$ and $h_4$, which describe the LOSVD's asymmetric and symmetric
deviations from a Gaussian. High signal-to-noise (S/N) spectra,
typically $\gtrsim 30$, are required in order to reliably extract the
higher order GH moments \citep[e.g.,][]{vanderMarel_Franx_1993,
  Bender_1994}, and thus we use the Voronoi binning algorithm
\citep{Cappellari_Copin_2003} in order to construct spatial bins that
optimize the balance between spatial resolution and S/N. We then used
the penalized pixel fitting (pPXF) method of
\cite{Cappellari_Emsellem_2004} to measure the stellar kinematics in
each bin. This procedure determines the best-fitting LOSVD by
convolving with a stellar template to match the observed galaxy
spectra. Errors on the kinematics were determined using Monte Carlo
simulations, in which random Gaussian noise was added to the spectrum
based upon the pPXF model residuals. We performed 100 realizations and
from the distributions measured the standard deviation to determine
1$\sigma$ uncertainties. During the Monte Carlo runs, the penalization
term was set to zero to produce realistic errors.

\subsection{NIFS Kinematics}
\label{subsec:nifskin}

We measured the stellar kinematics from the three primary $K$-band CO
bandheads [$(2-0)^{12}$CO, $(2-1)^{12}$CO, and $(4-2)^{12}$CO] in 127
spatial bins by using pPXF to fit the wavelength region between $2.26
- 2.42\ \mu$m. We made use of the NIFS Spectral Template Library
v2.0\footnote[9]{http://www.gemini.edu/sciops/instruments/nearir-resources/spectral-templates}
\citep{Winge_2009}, which contains 28 stars observed using the NIFS
IFU with the $K$ grating and $H + K$ filter. The library includes
spectral types ranging from G8 - M5 giant stars, K3 - M3 supergiants,
and a G8 II star.

We first created an optimal stellar template by fitting a high S/N
spectrum, constructed by adding together all spectra in the galaxy
data cube. This optimal template was composed of six stars, and was
dominated by M5 III, M3 III, and K3 Iab stars that make up 37\%, 26\%,
and 21\% of the total flux, respectively. Next, we measured the GH
moments in each spatial bin with pPXF by keeping the relative weights
of the stars that make up the template fixed, but allowing the
coefficients of a second degree additive Legendre polynomial and a
second degree multiplicative Legendre polynomial to vary. Such
polynomials are needed to account for differences in the optimal
stellar template and the galaxy spectra shape, as the continuum of the
stars in the NIFS Spectral Template Library has been previously
removed. The kinematics were in good agreement with those measured
when fitting a new optimal stellar template to each spatial bin and
fitting only two GH moments. Finally, we bi-symmetrize the kinematics
using the machinery presented \cite{vandenBosch_deZeeuw_2010}. Since
the dynamical models are only able to produce symmetric kinematics,
this step is commonly performed \citep[e.g.,][]{Gebhardt_2003,
  Cappellari_2006, Onken_2014} in order to reduce the noise in the
observations. With the symmetrization routine, the systematic offsets
in the odd GH moments, such as the galaxy's recession velocity, are
removed as well.

\begin{figure}
\begin{center}
\epsscale{1.0}
\plotone{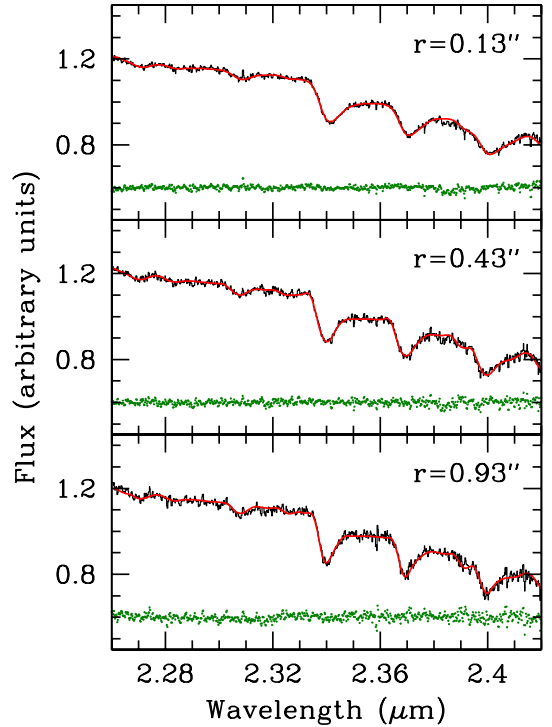}
\caption{Shown in the top, middle, and bottom panels, respectively,
  are example NIFS spectra extracted from three spatial locations: a
  single spaxel located near the nucleus, a bin containing five
  spaxels at an intermediate distance from the galaxy center, and one
  of the outermost bins composed of 28 spaxels. Overplotted in red is
  the optimal stellar template convolved with the best-fitting
  LOSVD. The model residuals are shown in green, and have been shifted
  by an arbitrary amount. \label{fig:spec_nifs}}
\end{center}
\end{figure}

From the NIFS data, we find that the galaxy is rotating quickly, with
stars reaching velocities of $\pm226$ km s$^{-1}$. There is also a
sharp peak in the velocity dispersion, which rises from 205 km
s$^{-1}$ at a radius of $\sim$1\arcsec\ to 396 km s$^{-1}$ at the
center. The map of $h_3$ is anti-correlated with the velocity map,
while $h_4$ shows a slight increase at the center. The S/N in each
spatial bin (measured as the ratio between the median value of the
spectrum and standard deviation of the pPXF model residuals) ranged
between 33 and 96 with a median value of 66. Therefore, we were able
to place excellent constraints on the kinematics, with median errors
over all spatial bins of $7$ km s$^{-1}$, $9$ km s$^{-1}$, $0.02$, and
$0.02$ for $V$, $\sigma$, $h_3$, and $h_4$, respectively. We present
example spectra and fits with pPXF at three different locations within
the NIFS data cube in Figure \ref{fig:spec_nifs} and the
bi-symmetrized NIFS kinematics in Table \ref{tab:nifskin}.

\begin{deluxetable*}{cccccccccc}
\tabletypesize{\scriptsize}
\tablewidth{0pt}
\tablecaption{NIFS Kinematics \label{tab:nifskin}}
\tablehead{
\colhead{$x$ (\arcsec)} &
\colhead{$y$ (\arcsec)} &
\colhead{$V$ (km s$^{-1}$)} &
\colhead{$\Delta V$ (km s$^{-1}$)} &
\colhead{$\sigma$ (km s$^{-1}$)} &
\colhead{$\Delta \sigma$ (km s$^{-1}$)} &
\colhead{$h_3$} &
\colhead{$\Delta h_3$} &
\colhead{$h_4$} &
\colhead{$\Delta h_4$} \\
\colhead{(1)} &
\colhead{(2)} &
\colhead{(3)} &
\colhead{(4)} &
\colhead{(5)} &
\colhead{(6)} &
\colhead{(7)} &
\colhead{(8)} &
\colhead{(9)} &
\colhead{(10)}
}

\startdata

  -0.027  &   0.022  &    -8.879  &   7.817  &   396.368  &  10.793  &   0.017  &   0.016  &   0.029  &  0.017  \\
  -0.027  &  -0.028  &   -38.385  &   6.554  &   394.762  &   8.844  &   0.017  &   0.014  &   0.033  &  0.018  \\
   0.055  &   0.022  &    37.125  &   5.608  &   395.394  &   7.818  &  -0.015  &   0.012  &   0.031  &  0.015  \\
   0.055  &  -0.028  &    11.045  &   6.073  &   392.972  &   8.368  &  -0.001  &   0.012  &   0.029  &  0.014  \\
  -0.077  &  -0.003  &   -45.778  &   6.763  &   392.601  &   9.207  &  0.026  &   0.014  &   0.028  &  0.017  \\

\enddata

\tablecomments{Table \ref{tab:nifskin} is published in its entirety in
  the electronic edition of ApJ. A portion is shown here for guidance
  regarding its form and content. Columns (1) and (2) are the $x$ and
  $y$ Voronoi bin generators, measured relative to the galaxy
  center. Columns (3) - (8) provide the bi-symmetrized NIFS kinematics
  and errors. The position angle is 141.14$^\circ$, measured
  counter-clockwise from the galaxy's major axis to $x$.}

\end{deluxetable*}

\subsection{PPAK Kinematics}
\label{subsec:ppakkin}

We measured the stellar kinematics in 268 spatial bins over a
wavelength range of 4200$-$7000 \AA, which includes a number of
absorption features such as the H$\beta$, Mg $Ib$, and Fe 5015
lines. During the fit with pPXF, we masked sky features and emission
lines and included a 15th degree additive polynomial. The kinematics
were extracted using the Indo-U.S. Library of Coud\'{e} Feed Stellar
Spectra \citep{Valdes_2004}, and the optimal stellar template was
dominated by G9 V, G9 III, K0 III, and A0p stars. As a final step, we
bi-symmetrize the kinematics and subtract off the systematic offsets
in the odd GH moments using the procedure in
\cite{vandenBosch_deZeeuw_2010}.

\begin{figure*}
\begin{center}
\epsscale{0.7}
\plotone{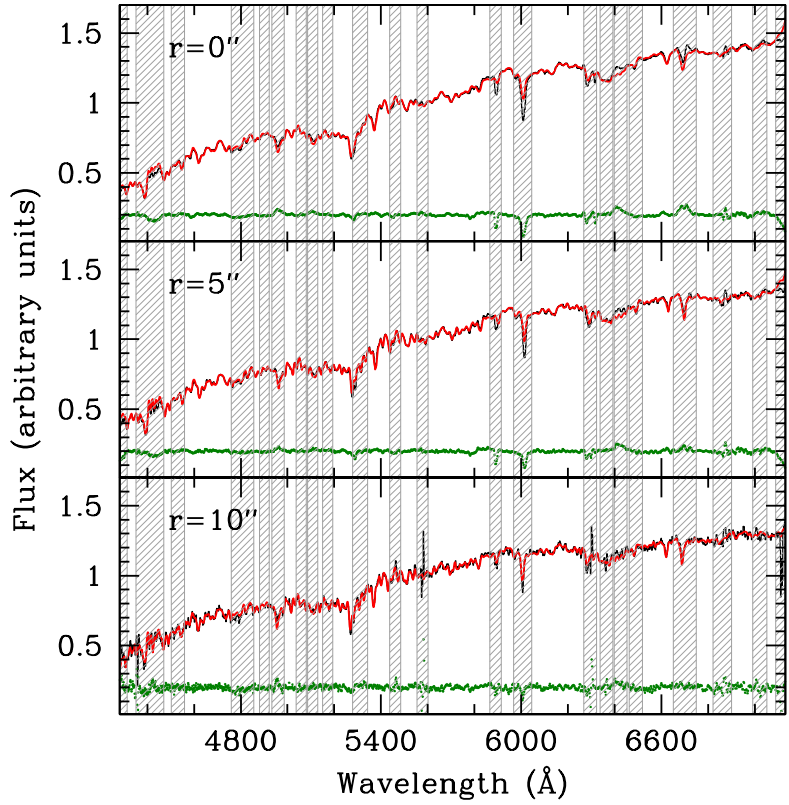}
\caption{Shown in the top, middle, and bottom panels, respectively,
  are example PPAK spectra extracted from three spatial
  locations. Overplotted in red is the optimal stellar template
  convolved with the best-fitting LOSVD, and the gray shaded boxes
  denote the wavelength regions excluded during the spectral fitting,
  due to the presence of emission lines or sky lines. The model
  residuals are shown in green, and have been shifted by an arbitrary
  amount. \label{fig:specfit_ppak}}
\end{center}
\end{figure*}

The kinematics from the PPAK data were measured out to
$\sim$24\arcsec, or $\sim$4 $R_e$. The large-scale kinematics exhibit
features that are similar to the measurements made from the high
spatial resolution NIFS data. In particular, the stars show rotation
with velocities of $\pm$231, a peak in the velocity dispersion to
values of 297 km s$^{-1}$ from 102 km s$^{-1}$ at radius of
$\sim$24\arcsec, and there is a $h_3 - V$ anti-correlation. The
difference in the peak velocity dispersions measured from the PPAK and
NIFS data can be attributed to the very different spatial resolutions
of the two data sets. Typical errors on the PPAK kinematics for $V$,
$\sigma$, $h_3$, and $h_4$ are 8 km s$^{-1}$, 12 km s$^{-1}$, 0.04,
and 0.05. We present example fits to the galaxy spectra at several
spatial locations in Figure \ref{fig:specfit_ppak} and provide the
bi-symmetrized PPAK kinematics in Table \ref{tab:ppakkin}.

NGC 1271 has an uncertain SB0 classification according to NED, but we
do not see an obvious bar feature in the \emph{HST} F160W image or in
the PPAK/NIFS kinematics. N-body simulations have shown that common
kinematic signatures associated with bars include a ``double-hump''
feature in the rotation curve and an $h_3 - V$ correlation over the
projected length of the bar \citep{Bureau_Athanassoula_2005}. Instead,
even the unsymmetrized kinematics clearly show a smooth increase in
the radial velocity from the southeast side of the galaxy to the
northwest side, and that $h_3$ is anti-correlated with $V$, as is
expected for axisymmetric systems.

\begin{deluxetable*}{cccccccccc}
\tabletypesize{\scriptsize}
\tablewidth{0pt}
\tablecaption{PPAK Kinematics \label{tab:ppakkin}}
\tablehead{
\colhead{$x$ (\arcsec)} &
\colhead{$y$ (\arcsec)} &
\colhead{$V$ (km s$^{-1}$)} &
\colhead{$\Delta V$ (km s$^{-1}$)} &
\colhead{$\sigma$ (km s$^{-1}$)} &
\colhead{$\Delta \sigma$ (km s$^{-1}$)} &
\colhead{$h_3$} &
\colhead{$\Delta h_3$} &
\colhead{$h_4$} &
\colhead{$\Delta h_4$} \\
\colhead{(1)} &
\colhead{(2)} &
\colhead{(3)} &
\colhead{(4)} &
\colhead{(5)} &
\colhead{(6)} &
\colhead{(7)} &
\colhead{(8)} &
\colhead{(9)} &
\colhead{(10)}
}

\startdata

  -0.063  &  -0.208  &   -28.540  &   4.759  &   296.999  &   6.723  &   0.010  &   0.013  &   0.027  &  0.016 \\
  -0.063  &   0.792  &    18.003  &   4.564  &   294.131  &   6.221  &  -0.013  &   0.013  &   0.025  &  0.016 \\
   0.937  &  -0.208  &    60.969  &   4.491  &   282.870  &   5.858  &  -0.030  &   0.013  &   0.035  &  0.017 \\
  -1.063  &  -0.208  &   -85.825  &   4.277  &   278.884  &   5.984  &   0.043  &   0.012  &   0.036  &  0.017 \\
  -0.063  &  -1.208  &   -60.969  &   4.491  &   282.870  &   5.858  &  0.030  &   0.013  &   0.035  &  0.017 \\

\enddata

\tablecomments{Table \ref{tab:ppakkin} is published in its entirety in
  the electronic edition of ApJ. A portion is shown here for guidance
  regarding its form and content. Columns (1) and (2) are the $x$ and
  $y$ Voronoi bin generators, measured relative to the galaxy
  center. Columns (3) - (8) provide the bi-symmetrized PPAK kinematics
  and errors. The position angle is 140.71$^\circ$, measured
  counter-clockwise from the galaxy's major axis to $x$.}

\end{deluxetable*}

\section{Measuring the PSF}
\label{sec:psf}

The PSF of the NIFS and PPAK observations are important inputs into
the stellar dynamical models. In order to estimate these quantities,
we convolve the MGE model presented in Section
\ref{sec:stellarmassmodel} with the sum of two concentric, circular 2D
Gaussians in order to match the collapsed NIFS and PPAK data
cubes. The PSF is parameterized by the dispersion and relative weight
of each Gaussian component. We find dispersions of 0\farcs16 and
0\farcs43 with relative weights of 0.61 and 0.39, respectively, for
the NIFS PSF, while the PPAK PSF can be described with dispersions of
1\farcs52 and 5\farcs45 with relative weights of 0.82 and 0.18. In
addition, the comparison between the \emph{HST} image and the
collapsed data cubes allows for the center of the NIFS and PPAK
apertures to be defined. The core of the NIFS PSF is larger than
expected for AO observations (e.g., \citealt{Krajnovic_2009,
  Seth_2014}), which is likely the result of using a fairly faint,
off-axis tip-tilt star. In Section \ref{subsec:errorbudget}, we test
the effect of our assumed NIFS PSF on the inferred black hole mass by
instead estimating the PSF from NIFS observations we acquired of the
tip-tilt star itself.

\section{Orbit-based Models}
\label{sec:models}

Black hole masses are often measured from stellar kinematics by
constructing dynamical models based upon the Schwarzschild
superposition method \citep{Schwarzschild_1979}, and here we use the
three-integral, triaxial Schwarzschild code of
\cite{vandenBosch_2008}. This technique finds a self-consistent
distribution function from the observables without any assumptions
about the orbital anisotropy. In the model, the black hole, the stars,
and dark matter all contribute to the galaxy's gravitational
potential. The stellar potential is determined by deprojecting the
observed surface brightness assuming a viewing orientation and a
stellar mass-to-light ratio ($\Upsilon$) that is constant with
radius. We then generate a representative orbit library in the
potential, and the orbits are numerically integrated while keeping
track of their intrinsic and projected properties. During the
modeling, the effects of the PSF and aperture binning are taken into
account. Finally, we assign weights to each orbit such that the
superposition matches the observed kinematics and total light
distribution. We calculate many models varying the parameters of
interest ($M_\mathrm{BH}$, $\Upsilon$, the viewing orientation
parameters, and the dark matter halo parameters), and the best-fit
model is the one with the lowest $\chi^2$.

\subsection{Application to NGC 1271}
\label{subsec:app_to_n1271}

For NGC 1271, we adopt a (nearly) oblate axisymmetric shape, with an
intermediate to long axis ratio of 0.99. We assume axisymmetry given
that NGC 1271 looks highly flattened and exhibits rapid rotation, and
we do not find any evidence for kinematic twists. With this
assumption, the inclination angle ($i$) is the only viewing
orientation parameter needed to describe the galaxy's intrinsic
shape. In our final model, we adopt $i=83^\circ$ (where $i=90^\circ$
corresponds to an edge-on view). The MGE description of NGC 1271
presented in Section \ref{sec:stellarmassmodel} includes a couple of
flat Gaussian components, which can only be deprojected for angles $77
\lesssim i \lesssim 90^\circ$. Therefore, we chose to run models at a
fixed inclination angle that lies midway between the extremes. Our
choice of inclination angle is further supported by a nuclear dust
disk that appears highly inclined in an \emph{HST} WFC3 F814W
image. While the F814W image clearly shows a regular dust lane, the
F160W image and the $K$-band NIFS data cube used in our analysis do
not appear to be significantly affected by dust. In addition, the NGC
1271 models include a spherically symmetric dark matter halo following
a Navarro-Frenk-White (NFW) form \citep{Navarro_1996}. The parameters
describing the NFW halo are the concentration index ($c$) and the
fraction of dark matter ($f_\mathrm{DM}$), where $f_\mathrm{DM} \equiv
M_\mathrm{DM}/M_\star$ and $M_\mathrm{DM}$ is the halo virial mass and
$M_\star$ is the stellar mass. Hence, our models have four free
parameters: $M_\mathrm{BH}$, $\Upsilon_{H}$, $c$, and $f_\mathrm{DM}$.
 
We began by calculating models on a coarse grid that spanned a wide
range of values before generating models on a smaller, more finely
sampled grid focused around the minimum $\chi^2$. Ultimately, our
final model grid contained 21 $M_\mathrm{BH}$ values, 31
$\Upsilon_{H}$ values, 8 $c$ values, and 20 $f_\mathrm{DM}$ values
with $\log(M_\mathrm{BH}/M_\odot) \in [9.2,10.2]$, $\Upsilon_{H}\
(\Upsilon_\odot) \in [0.5,2.0]$, $c \in [2,16]$, and $f_\mathrm{DM}
\in [50,1000]$. The orbit library samples 29 equipotential shells with
radii between 0\farcs003 to 100\arcsec, and 8 angular and 8 radial
values at each energy. We note that triaxial orbit families (e.g., box
orbits) continue to be included in our orbital libraries because we
are running a triaxial Schwarzschild code in the axisymmetric
limit. Moreover, we employ a dithering method, in which 125 orbits
with adjacent initial conditions are bundled together, to establish a
smooth distribution function when constructing Schwarzschild
models. Thus, the galaxy models are made with 696,000 orbits. The
models were fit to the observed NIFS and PPAK kinematics, where 4 GH
moments were measured in a total of 395 bins, resulting in 1580
observables.

\section{Modeling Results}
\label{sec:results}

\begin{figure*}
\begin{center}
\epsscale{0.9}
\plotone{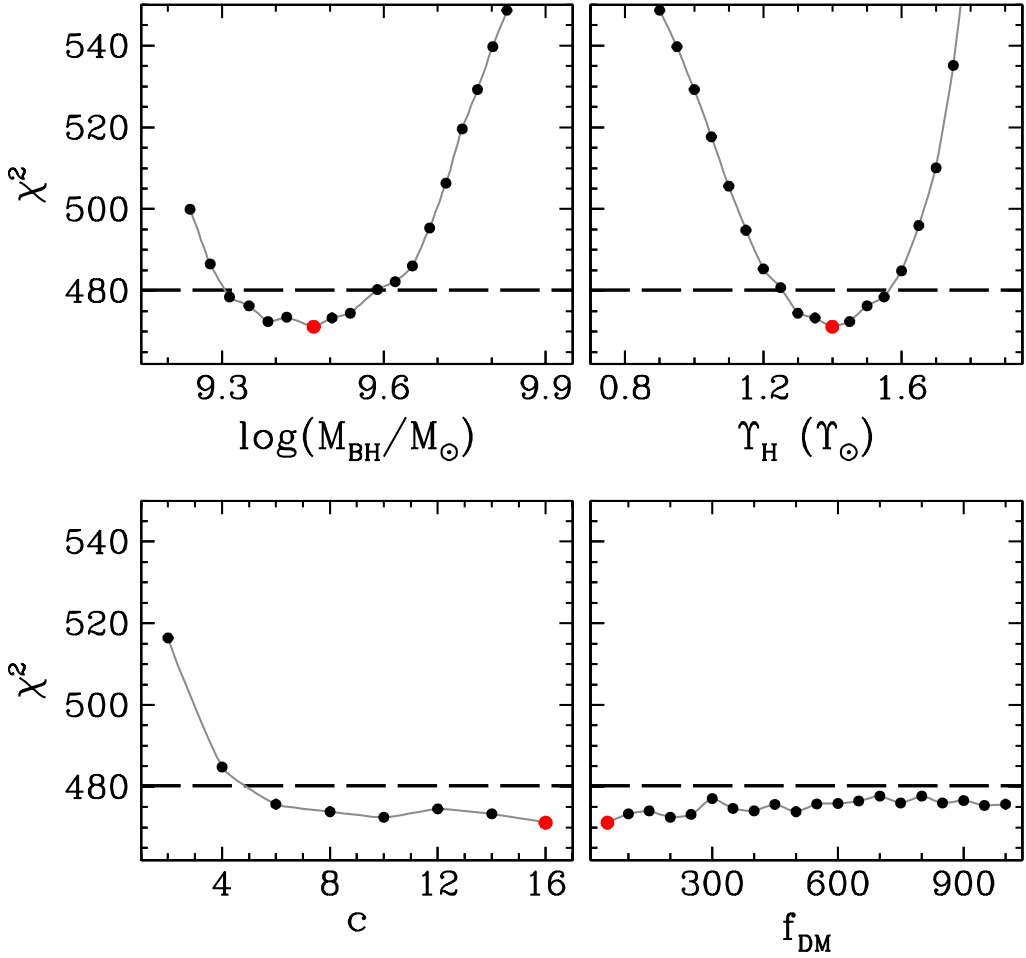}
\caption{The plots summarize the results of the stellar dynamical
  models run with $i= 83^\circ$ and an NFW dark matter halo. The
  $\chi^2$ is shown as a function of $M_\mathrm{BH}$ (top left),
  $\Upsilon_{H}$ (top right), $c$ (bottom right), and $f_\mathrm{DM}$
  (bottom left) after marginalizing over the other three
  parameters. The red point denotes the best-fit model, and the dashed
  line depicts where minimum $\chi^2$ has increased by 9, which
  corresponds to the statistical 3$\sigma$ uncertainties. The
  uncertainty in $M_\mathrm{BH}$ and $\Upsilon_H$ due to systematic
  effects is significantly larger than that statistical 1$\sigma$
  error (see Section \ref{subsec:errorbudget}). Thus, we do not plot
  the statistical 1$\sigma$ confidence level, as it is not
  representative of the range of possible parameter values. The
  $M_\mathrm{BH}$ and $\Upsilon_{H}$ parameters are well constrained
  despite the large uncertainties associated with the dark halo
  parameters. \label{fig:chi2plots}}
\end{center}
\end{figure*}

\begin{figure*}
\begin{center}
\epsscale{0.9}
\plotone{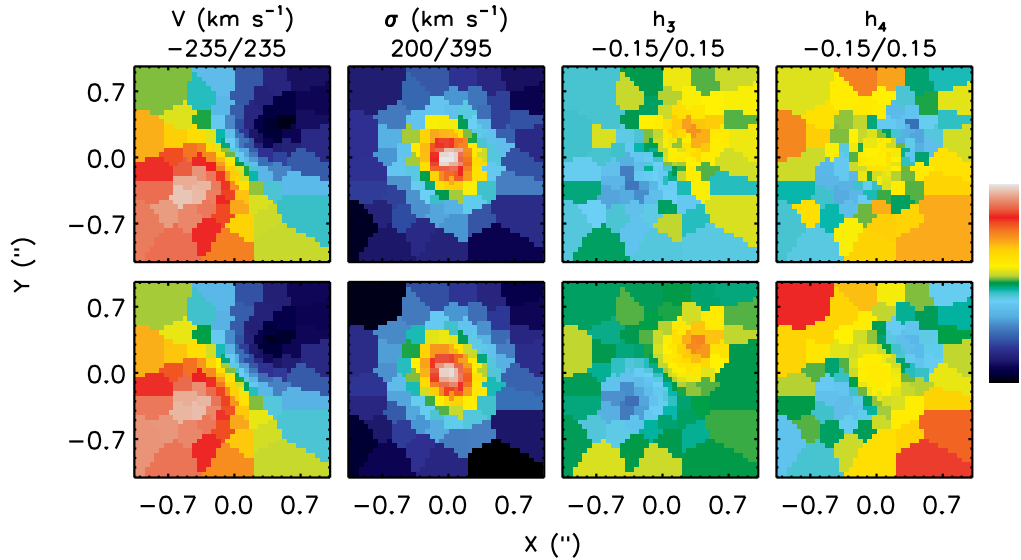}
\caption{The bi-symmetrized NIFS kinematics for NGC 1271 (top) are
  compared to the best-fit model predictions (bottom), where
  $M_\mathrm{BH} = 3.0\times10^9\ M_\odot$ and $\Upsilon_{H} = 1.40\
  \Upsilon_\odot$. The same scaling, shown by the color bar on the
  right with the minimum and maximum values given at the top of the
  maps, is used to plot the data and model. The NIFS observations show
  that the galaxy is rapidly rotating with a peak in the velocity
  dispersion at the nucleus. An anti-correlation between $h_3$ and $V$
  is found, as is expected for galaxies with axial symmetry. The
  blue-shifted side of the radial velocity map corresponds to the
  southeast side of the galaxy. \label{fig:nifs_datamodel}}
\end{center}
\end{figure*}

\begin{figure*}
\begin{center}
\epsscale{1.1}
\plotone{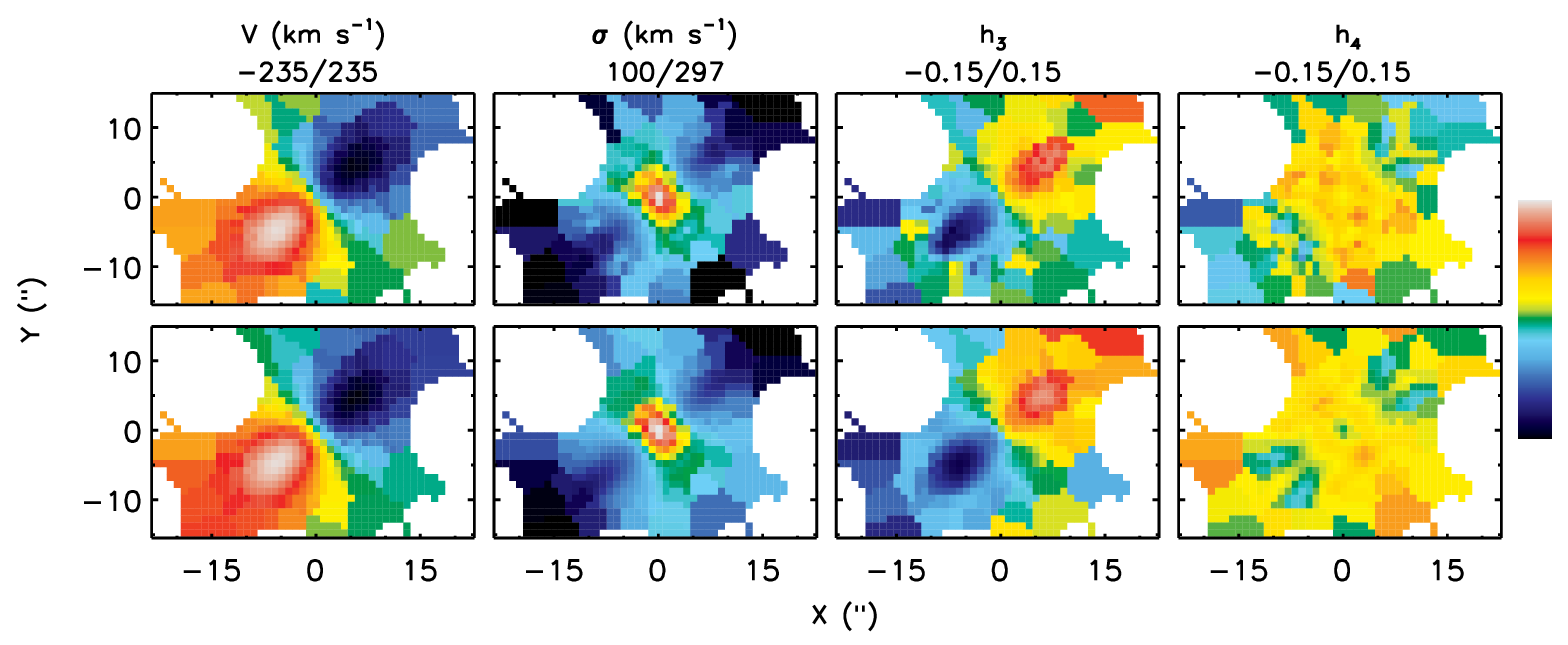}
\caption{The bi-symmetrized PPAK kinematics (top), which extend out to
  about 4 $R_e$, are shown along with the predictions from the
  best-fit stellar dynamical model (bottom); see Figure
  \ref{fig:nifs_datamodel} for description. Kinematic measurements are
  missing from $x$$\sim$$-15$\arcsec, $y$$\sim$$10$\arcsec\ and from
  $x$$\sim$$20$\arcsec, $y$$\sim$$0$\arcsec\ due to the presence of
  foreground objects, which were masked before extracting the
  kinematics. The southeast side of the galaxy has blue-shifted radial
  velocities. \label{fig:ppak_datamodel}}
\end{center}
\end{figure*}

The results of the final model grid are summarized in Figure
\ref{fig:chi2plots}, which shows the $\chi^2$ as a function of
$M_\mathrm{BH}$, $\Upsilon_{H}$, $c$, and $f_\mathrm{DM}$ after
marginalizing over the other three parameters. We find best-fit values
of $M_\mathrm{BH} = 3.0\times10^9\ M_\odot$, $\Upsilon_{H} = 1.40\
\Upsilon_\odot$, $c = 16$, and $f_\mathrm{DM} = 50$ (corresponding to
$M_\mathrm{DM} = 5.0\times10^{12}\ M_\odot$). The comparison between
the observed NIFS kinematics and model predictions is shown in Figure
\ref{fig:nifs_datamodel} and the comparison between the PPAK data and
model predictions is displayed in Figure \ref{fig:ppak_datamodel}. The
model is an excellent match to the observed kinematics, and is able to
reproduce the sharp increase in the velocity dispersion and the slight
peak in $h_4$ at the nucleus. The $\chi^2$ per degree of freedom
($\chi^2_\nu$) is 0.3. If the unsymmetrized kinematics are used
instead, the best-fit model has $\chi^2_\nu = 1.3$ and an
$M_\mathrm{BH}$ and $\Upsilon_H$ that are consistent within the final
uncertainties given below in Section \ref{subsec:errorbudget}.

As can be seen in Figure \ref{fig:chi2plots}, we are able to place
strong constraints on $M_\mathrm{BH}$ and $\Upsilon_{H}$. By searching
for the range of $M_\mathrm{BH}$, or $\Upsilon_{H}$, values that
caused the minimum $\chi^2$ to increase by 1 and 9, we estimate the
1$\sigma$ and 3$\sigma$ model fitting uncertainties. We find
$M_\mathrm{BH} = (3.0^{+0.1}_{-0.6}) \times 10^9\ M_\odot$ (1$\sigma$)
and $M_\mathrm{BH} = (3.0^{+0.9}_{-1.0}) \times 10^9\ M_\odot$
(3$\sigma$), as well as $\Upsilon_{H} = 1.40^{+0.05}_{-0.02}\
\Upsilon_\odot$ (1$\sigma$) and $\Upsilon_{H} = 1.40^{+0.16}_{-0.15}\
\Upsilon_\odot$ (3$\sigma$). In contrast, the dark halo parameters are
not well constrained, and the $c$ and $f_\mathrm{DM}$ values are
highly uncertain. Specifically, the marginalized $\chi^2$ curve for
$c$ is unconstrained at the upper end, with $c > 5$ at the 3$\sigma$
level. Although the best-fit value for $c$ is at the maximum value
considered in our final model grid, previous runs using coarsely
sampled grids showed no sign of convergence for large values of
$c$. Similarly, the marginalized $\chi^2$ curve for $f_\mathrm{DM}$ is
essentially flat, and all $f_\mathrm{DM}$ values sampled by our grid
($50 \leq f_\mathrm{DM} \leq 1000$) are allowed within 3$\sigma$. The
best-fit value found for $f_\mathrm{DM}$ is the minimum value
considered in our final grid, but we also ran models without a dark
halo and found a significantly worse fit, such that the $\chi^2$
increased by 123 relative to the best-fit model with a dark
halo. Thus, a dark halo is required to match the observed kinematics,
but its properties cannot be pinned down with the current datasets and
we do not present any further details associated with the dark halo
parameters.

\subsection{Error Budget}
\label{subsec:errorbudget}

The formal model fitting uncertainties quoted in the previous section
are the statistical errors associated with the dynamical models,
however systematic effects can have a significant impact on the
inferred black hole mass and mass-to-light ratio. In this section, we
evaluate some common sources of uncertainty that are not already
incorporated into the statistical errors, such as those associated
with the assumed form of the dark matter halo, the adopted inclination
angle, the number of orbits used in the models, the NIFS PSF model,
details associated with the extraction of the NIFS kinematics, and the
symmetrization of the input kinematics.

\emph{Dark Matter Halo}: Previous work has clearly shown that the dark
halo could be an important component in the stellar dynamical models
due to the degeneracy between the dark halo and stellar mass-to-light
ratio, which in turn is also degenerate with the black hole
\citep[e.g.,][]{Gebhardt_Thomas_2009, Schulze_Gebhardt_2011,
  Rusli_2013}. If $r_\mathrm{sphere}$ is spatially well resolved, then
the degeneracy between the black hole and mass-to-light ratio can be
mitigated, and the exclusion of a dark halo will have minimal impact
on the inferred $M_\mathrm{BH}$. In contrast, if $r_\mathrm{sphere}$
is not very well resolved and a dark halo is not included in the
stellar dynamical models, then $\Upsilon$ will be artificially
elevated to account for the missing mass at large radii. Since
$\Upsilon$ is taken to be constant with radius, a smaller
$M_\mathrm{BH}$ is then required to fit the observed central
kinematics.

When generating a grid of models without a dark halo for NGC 1271, and
assuming $i=83^\circ$ to match our fiducial model presented in Section
\ref{sec:results}, we measure $M_\mathrm{BH} = 1.4 \times 10^9\
M_\odot$ and $\Upsilon_{H} = 1.85\ \Upsilon_\odot$. In other words,
the inferred $M_\mathrm{BH}$ is underestimated by a factor of $\sim$2
when a dark halo is excluded from the modeling. We note that when
fitting dynamical models without a dark halo to only the small-scale
NIFS kinematics, we recover very similar parameter values as those
found from our fiducial model, with $M_\mathrm{BH} = 2.9\times10^9\
M_\odot$ and $\Upsilon_H = 1.45\ \Upsilon_\odot$. Moreover, we tested
how the assumed shape of the dark halo affects $M_\mathrm{BH}$. Our
fiducial model was calculated assuming a spherical NFW halo, but
another common form is a halo with a cored logarithmic potential
\citep{Binney_Tremaine_1987, Thomas_2005}, given by

\begin{equation}
\label{eq:logdm}
\rho_\mathrm{DM}(r) = \frac{V_c^2}{4 \pi G} \frac{3r_c^2 + r^2}{(r_c^2
  + r^2)^2} .
\end{equation}

\noindent The parameters $V_c$ and $r_c$ are the asymptotic circular
velocity and radius within which the dark matter density is
constant. Thus, the halo from a logarithmic potential yields smaller
densities at small radii compared the the NFW halo. We constructed
models with a dark halo from a cored logarithmic potential, sampling
$9.2 \leq \log(M_\mathrm{BH}/M_\odot) \leq 10.2$, $0.5 \leq
\Upsilon_{H} \leq 2.0$ $\Upsilon_\odot$, $100 \leq V_c \leq 700$ km
s$^{-1}$, and $1 \leq r_c \leq 32$ kpc. We recovered similar results
to the models with an NFW halo, namely that $M_\mathrm{BH} = 3.2
\times 10^9\ M_\odot$ and $\Upsilon_H = 1.35\ \Upsilon_\odot$,
corresponding to a 7\% increase in $M_\mathrm{BH}$ and a 4\% decrease
in $\Upsilon_H$ compared to the fiducial model. Although the black
hole mass is sensitive to the inclusion of a dark halo in the stellar
dynamical models, the form of the halo has a small impact on
$M_\mathrm{BH}$. For this reason, being unable to constrain the dark
halo parameters is not a concern for the purposes of this paper, as
long as reasonable halos are sampled over when constructing the
orbit-based models.

\emph{Inclination Angle}: All of the models presented in this paper
assume an axisymmetric shape with an inclination angle of
$i=83^\circ$. However, we also ran a grid of models that sampled 13
inclination angles from $77$ to $89^\circ$. This corresponds to the
range of angles for which the MGE model in Section
\ref{sec:stellarmassmodel} can be deprojected. It is computationally
expensive to calculate a model grid that samples over $M_\mathrm{BH}$,
$\Upsilon_H$, $i$, $c$, and $f_\mathrm{DM}$ simultaneously, so we
instead varied the first three parameters while sampling over five NFW
halos. The five dark halos were those with the lowest $\chi^2$ from
the model grid at the beginning of Section \ref{sec:results}, and the
halos span a range of $c$ and $f_\mathrm{DM}$ values ($8 \leq c \leq
16$ and $50 \leq f_\mathrm{DM} \leq 500$). From this test, we
determine that $M_\mathrm{BH} = 3.3 \times 10^9\ M_\odot$, which is
within 10\% of the best-fit value in Section \ref{sec:results}, and
$\Upsilon_H = 1.40\ \Upsilon_\odot$, which is the same as the best-fit
value in Section \ref{sec:results}. Moreover, the best-fit inclination
angle was $i=87^\circ$, however the angle was not well
constrained. All angles between $79^\circ$ and $89^\circ$ were allowed
within the 3$\sigma$ statistical uncertainties. Such behavior is not
surprising, and other stellar dynamical work have also found it
difficult to infer the inclination angle from 2D line-of-sight
kinematics \citep{Krajnovic_2005, vandenBosch_vandeVen_2009}.

\emph{Number of Orbits}: The fiducial model presented in Section
\ref{sec:results} was calculated using orbits that covered 29
equipotential shells with 8 angular and 8 radial values at each
energy. When accounting for the orbital dithering, this translates
into a total of 696,000 orbits. We also tested the effect on
$M_\mathrm{BH}$ and $\Upsilon_H$ when the number of orbits is about
doubled, such that 37 equipotential shells with 10 angular and 10
radial values at each energy are used. Again 125 orbits with adjacent
starting positions were bundled together, resulting in 1,387,500
orbits. Due to a large increase in computational time for a single
model, we constructed a model grid that samples over $M_\mathrm{BH}$,
$\Upsilon_H$, and the top five NFW halos from Section
\ref{sec:results}. We found no change in the best-fit $M_\mathrm{BH}$
or $\Upsilon$ values compared to the fiducial model.

\emph{NIFS PSF}: The NIFS PSF was measured by comparing the MGE model
of the \emph{HST} image to the collapsed NIFS data cube. While this
approach is commonly used in black hole mass measurement work
\citep[e.g.,][]{Krajnovic_2009, Seth_2010, Walsh_2012}, estimation of
the PSF from AO observations is notoriously difficult due to the
constantly changing quality of the AO correction and the combination
of data cubes from multiple nights. Therefore, we also estimated the
PSF in a different manner in order to assess how strongly the adopted
NIFS PSF affects $M_\mathrm{BH}$. Utilizing the NIFS observations of
the tip-tilt star, we fit the sum of four concentric, circular 2D
Gaussians to the collapsed NIFS data cube using Galfit. We found that
the PSF is best described by Gaussians with dispersions of 0\farcs04,
0\farcs08, 0\farcs21, and 0\farcs43 with relative weights of 0.08,
0.36, 0.26, and 0.30. A four-component PSF model provided a
significantly better fit to the collapsed data cube than a simpler
two-component model, and the residuals between the four Gaussian model
and the data have a standard deviation of just 9\% out to a radius of
1\arcsec. The PSF is typical of what one would expect from the Gemini
AO system \citep[e.g.,][]{Gebhardt_2011, Onken_2014}, and the quality
of the AO correction is better than that implied by the PSF adopted in
the fiducial model. This is not completely surprising as the NIFS
observations of the star used the star itself for tip-tilt
corrections, whereas the observations of the galaxy were made
off-axis, using the star for guiding. Nonetheless, calculating stellar
dynamical models using this better PSF and comparing to the results
using the poorer PSF should cover the range of possible black hole
masses due to the uncertainty in the NIFS PSF. We calculated models
using the new four-Gaussian PSF and further assumed that the center of
the NIFS spaxel with the largest flux coincides with the galaxy
nucleus. We varied $M_\mathrm{BH}$ and $\Upsilon_H$, while sampling
over the top five NFW halos from Section \ref{sec:results}, and found
$M_\mathrm{BH} = 2.7 \times 10^9\ M_\odot$ and $\Upsilon_H = 1.45\
\Upsilon_\odot$. Therefore, the black hole mass and mass-to-light
ratio change by 10\% and 4\% compared to the fiducial values in
Section \ref{sec:results}.

\emph{Measuring the NIFS Kinematics:} We measured the NIFS kinematics
with pPXF using a second degree additive Legendre polynomial and a
second degree multiplicative polynomial to correct for shape
differences between the LOSVD-broadened optimal stellar template and
the observed galaxy spectrum. We selected this continuum correction
because it was one of the simplest models that still provided a good
fit to the data, and produced kinematics that were in good agreement
with those measured using combinations of degree $0-3$
additive/multiplicative polynomials, with the exception of the lowest
order polynomials. Since the continuum has been previously divided out
of the the spectral templates in the NIFS library, but not removed
from the observed galaxy spectra, using the lowest order polynomials
(combinations of degree $0-1$ additive/multiplicative polynomials)
produced visibly poor spectral fits.

We also constructed a new library containing stellar templates that
have not been continuum-corrected and thus have a very similar shape
as the galaxy spectra. We retrieved $K$-band NIFS observations from
the Gemini archive of stars that are part of the NIFS Spectral
Template Library and two stars observed under program
GN-2010A-Q-112. These twelve stars are K0 - M5 giants, a K5
supergiant, and an M0 supergiant. We reduced the observations
following the main procedure outlined in Section \ref{subsec:obsnifs},
with the additional steps of extracting a one-dimensional spectrum,
rebinning to a common wavelength range and sampling, and shifting the
stars to rest. With this new template library, we are able to obtain
good fits to the galaxy spectra using low-order polynomials with pPXF.

In order to examine possible effects on $M_\mathrm{BH}$ and
$\Upsilon_H$ due to uncertainties associated with the choice of the
pPXF polynomial degree, we fit dynamical models to the NIFS kinematics
extracted using the new stellar template library and an additive
constant, along with the PPAK kinematics presented in Section
\ref{subsec:ppakkin}. We sampled over $M_\mathrm{BH}$, $\Upsilon_H$,
and the top five NFW halos from Section \ref{sec:results}, finding
$M_\mathrm{BH} = 2.4 \times10^9\ M_\odot$ and $\Upsilon_H = 1.45\
\Upsilon_\odot$. This corresponds to a change of 20\% and 4\% from the
best-fit black hole mass and mass-to-ratio from the fiducial
model. These results are likely representative of a maximum change in
best-fit parameter values, as using this particular continuum
correction with the new stellar template library resulted in the
largest number of bins with kinematics inconsistent at the 1$\sigma$
level (all bins were consistent at the 2$\sigma$ level) compared to
the NIFS kinematics from Section \ref{subsec:nifskin}.

\emph{Symmetrizing the Kinematics:} The NIFS and PPAK kinematics were
bi-symmetrized prior to using them as inputs into the stellar
dynamical modeling code. We used the method outlined in
\cite{vandenBosch_deZeeuw_2010}, which averages the measurements of a
GH moment in a four-fold symmetric manner around the minor and major
axes. During the averaging of the measurements for a single GH moment,
the kinematic error of the bin and the fraction of flux a spaxel
contributes to that bin are taken into account. Such modifications to
the input kinematics and their errors are a common way in which to
reduce observational noise, and often kinematics are bi-symmetrized in
order to obtain reasonable results from an axisymmetric modeling code
and are point-symmterized for use with a triaxial modeling
code. However, symmterization routines are never perfect, as discussed
in \cite{vandenBosch_deZeeuw_2010}. We therefore tested how
symmetrization affects $M_\mathrm{BH}$ and $\Upsilon_H$ by running
additional models with kinematics that were point-symmetrized. We find
that the black hole mass increases to $3.6\times10^9\ M_\odot$, or by
20\% of the fiducial value, and the $H$-band mass-to-light ratio
decreases to $1.35\ \Upsilon_\odot$, or 4\% of the fiducial value from
Section \ref{sec:results}.

\emph{Summary:} We derive the final range of black hole masses and
mass-to-light ratios for NGC 1271 by adding in quadrature the formal
model fitting 1$\sigma$ uncertainty from Section \ref{sec:results} and
the additional sources of systematic uncertainty above. Ultimately, we
determine that $M_\mathrm{BH} = (3.0^{+1.0}_{-1.1}) \times 10^9\
M_\odot$ and $\Upsilon_H = 1.40^{+0.13}_{-0.11}\ \Upsilon_\odot$. The
dominant source of systematics for NGC 1271 are those associated with
the continuum-correction model used to extract the NIFS kinematics and
the symmetrization of the kinematics. Both affect $M_\mathrm{BH}$ and
$\Upsilon_H$ at the 20\% and 4\% level, respectively.

\subsection{Other Considerations}
\label{subsec:otherconsiderations}

In addition to examining possible sources of systematic uncertainty
and incorporating the effects into the final error budget, we ran
other tests to assess the robustness of the NGC 1271 $M_\mathrm{BH}$
measurement. We describe these tests below.

\emph{Nuclear Dust Disk:} A dust disk is present at the center of NGC
1271, and is visible in the F814W WFC3 image. However, dust doesn't
appear to be significant in the near-infrared imaging and in the NIFS
data cube. We tested constructing a new MGE of the F160W image after
excluding the dust disk by using the F814W image as guide for
generating the mask. We ran dynamical models using the modified MGE,
following the procedure in Section \ref{subsec:app_to_n1271} and
sampling over the top five NFW halos from Section
\ref{sec:results}. We found best-fit values of $M_\mathrm{BH} =
3.2\times10^9\ M_\odot$ and $\Upsilon_H = 1.35\ \Upsilon_\odot$, which
is well within the final uncertainties adopted for NGC 1271 discussed
in Section \ref{subsec:errorbudget}.

\emph{Stellar Mass-to-Light Ratio Variation:} Our dynamical models
assume that $\Upsilon_H$ remains constant with radius. In order to
determine whether there is an obvious change in stellar population, we
generated an MGE of the F814W image following the methods described in
Section \ref{sec:stellarmassmodel}. During the fit, we masked out the
nuclear dust disk and foreground objects in the F814W image, and we
accounted for the PSF using a bright, isolated star in the image. From
the MGE models of the F814W and F160W images, we don't see evidence
for color gradients, finding that the color changes by at most 0.16
mag from 0\farcs2 to 24\arcsec.

\begin{figure}
\begin{center}
\epsscale{1.0}
\plotone{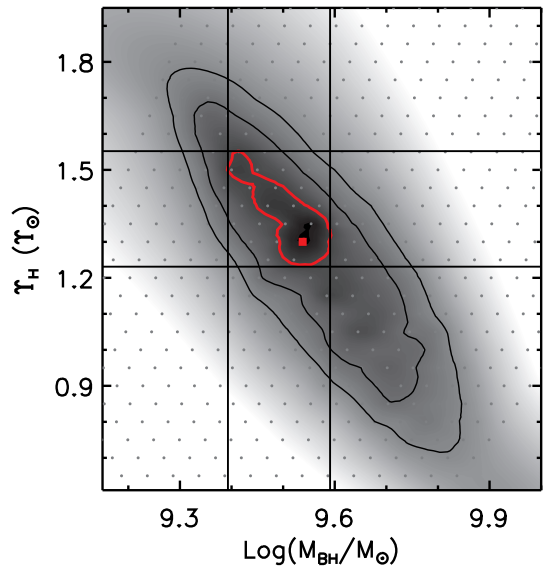}
\caption{Contours of $\chi^2$ are shown as a function of black hole
  mass and $H$-band mass-to-light ratio for the case when dynamical
  models are fit to just the NIFS kinematics. At each gray point a
  model was calculated, and the red square denotes the best-fit
  model. The red contour and two black contours signify where $\chi^2$
  has increased by 1, 4, and 9 from the minimum. Thus, the vertical
  lines show the 1$\sigma$ uncertainties on $M_\mathrm{BH}$ and the
  horizontal lines give the 1$\sigma$ uncertainties for $\Upsilon_H$
  when marginalizing over the other
  parameters. \label{fig:nifsonlygrid}}
\end{center}
\end{figure}

Although NGC 1271 exhibits a fairly uniform color, we further examined
dynamical models that are fit to only the NIFS kinematics. Given the
limited radial extent of the NIFS kinematics, which extend out to a
radius of $\sim$1\arcsec, or $\sim$390 pc, systematics associated with
mass-to-light ratio gradients (and dark matter halos) are
mitigated. When fitting to only the NIFS kinematics, we recover
consistent results to those presented in \ref{subsec:errorbudget},
where $M_\mathrm{BH} = (3.5^{+0.4}_{-1.0})\times10^9\ M_\odot$ and
$\Upsilon_H = 1.30^{+0.25}_{-0.07}\ \Upsilon_\odot$ (1$\sigma$
uncertainties). We show contours of $\chi^2$ as a function of black
hole mass and mass-to-light ratio for this NIFS-only model grid in
Figure \ref{fig:nifsonlygrid}.

\emph{PPAK Kinematics:} When measuring the stellar kinematics from the
PPAK data, we masked out spectral regions containing possible emission
lines and sky residuals, as can be seen in Figure
\ref{fig:specfit_ppak}. In order to verify that our choice of a
spectral mask does not bias the kinematic measurements and inferred
black hole mass, we decreased the number and width of the excluded
wavelength regions. We ran dynamical models using the modified PPAK
kinematics while sampling over the top five dark matter halos in
Section \ref{sec:results}. We measured $M_\mathrm{BH} = 3.2\times10^9\
M_\odot$ and $\Upsilon_H = 1.35\ \Upsilon_\odot$, which is within the
final uncertainties given for NGC 1271 in Section
\ref{subsec:errorbudget}.

Also, the observed velocity dispersion of the PPAK kinematics
presented in Section \ref{subsec:ppakkin} drops below the instrumental
resolution in many of the spatial bins located $\gtrsim$10\arcsec\
away from the nucleus. While care was taken to homogenize the line
spread function to a common value such that there was no variation
with wavelength or fiber position, measuring dispersions well below
the instrumental resolution is a difficult task. We therefore also
tested the effect on $M_\mathrm{BH}$ and $\Upsilon_H $ when excluding
the spatial bins in which the dispersion was below $150$ km
s$^{-1}$. When calculating dynamical models that fit to the adjusted
PPAK kinematics and that sample over the top five dark matter halos in
Section \ref{sec:results}, we find no change from the best-fit values
presented in Section \ref{sec:results}.

\section{Discussion}
\label{sec:discussion}

NGC 1271 harbors a black hole with $M_\mathrm{BH} =
(3.0^{+1.0}_{-1.1}) \times 10^9\ M_\odot$ and has a stellar
mass-to-light ratio of $\Upsilon_H = 1.40^{+0.13}_{-0.11}\
\Upsilon_\odot$. We note that the final uncertainty on the black hole
mass we use is comparable to the formal 3$\sigma$ statistical
uncertainty. Some \citep{Cappellari_2009, Krajnovic_2009,
  Emsellem_2013} have suggested 3$\sigma$ statistical errors should be
used in place of 1$\sigma$ errors as a conservative way in which to
account for the effect of unknown systematics on $M_\mathrm{BH}$. With
a black hole mass of $3.0\times10^9\ M_\odot$ and adopting 276 km
s$^{-1}$ for the bulge stellar velocity dispersion (see Section
\ref{subsec:bhrels}), $r_\mathrm{sphere} = $ 0\farcs44. Thus, the NIFS
observations have resolved the black hole sphere of influence. Below
we discuss the galaxy's orbital structure and place the galaxy on the
$M_\mathrm{BH}$$-$host galaxy relations.

\subsection{Orbital Structure}
\label{subsec:orbstructure}

\begin{figure}
\begin{center}
\epsscale{1.0}
\plotone{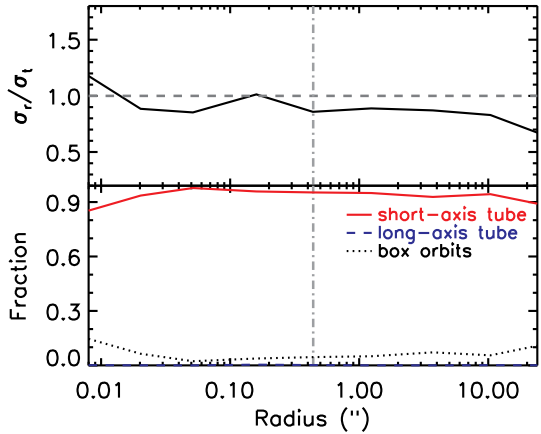}
\caption{NGC 1271's orbital structure, as inferred from the best-fit
  dynamical model, is shown. The anisotropy (top) and orbit type
  (bottom) are displayed with radius over the range covered by the
  NIFS and PPAK kinematic measurements. The horizontal dashed gray
  line denotes the isotropic case and the vertical dot-dashed gray
  line shows the black hole sphere of influence. NGC 1271 is roughly
  isotropic at all radii and is dominated by short-axis tube orbits as
  is expected for oblate systems. \label{fig:orbstructure}}
\end{center}
\end{figure}

In addition to determining the mass of the black hole in NGC 1271, the
Schwarzschild models provide information about the galaxy's orbital
structure. Using our best-fit model in Section \ref{sec:results}, we
show the ratio $\sigma_r/\sigma_t$ as a function of radius in Figure
\ref{fig:orbstructure}. The tangential velocity dispersion is defined
as $\sigma_t^2 = (\sigma_\phi^2 + \sigma_\theta^2)/2$, and $(r,
\theta, \phi)$ are the usual spherical coordinates. We find that NGC
1271 is roughly isotropic at all radii covered by our kinematic
measurements, deviating by at most 30\% from $\sigma_r/\sigma_t = 1$,
but we observe a trend in which $\sigma_r/\sigma_t$ declines at radii
outside the black hole sphere of influence. As expected, short-axis
tube orbits dominate in this oblate system, making up more than 85\%
of the orbits at all radii. Long-axis tube orbits, which are important
for triaxial and prolate systems, are negligible, while the fraction
of box orbits increases at small radii but still make up only 15\% of
the orbits near the nucleus.

\begin{figure}
\begin{center}
\epsscale{1.0}
\plotone{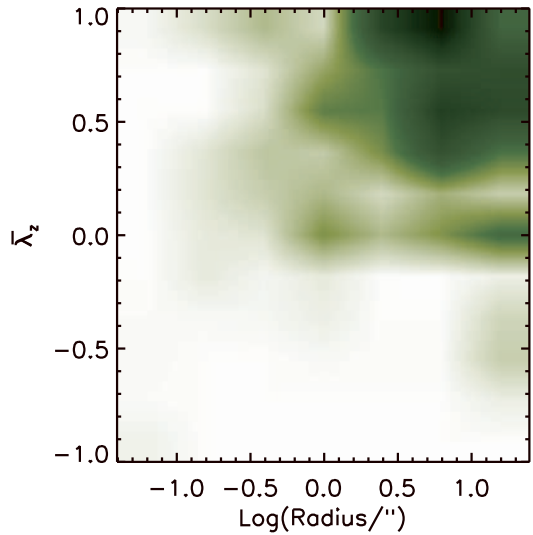}
\caption{The mass distribution is plotted as a function of average
  spin and radius of the orbits for the region covered by our
  kinematics measurements. The dynamical decomposition utilizes the
  orbital weights from our best-fit stellar dynamical model and shows
  distinct non-rotating bulge ($-0.2 < \bar \lambda_z < 0.2$) and
  rotating ($\bar \lambda_z > 0.2$) disk
  components.\label{fig:massdens}}
\end{center}
\end{figure}

Furthermore, we use our best-fit stellar dynamical model to examine
the mass distribution as a function of average radius, $\bar r$, and
spin, $\bar \lambda_z$, of the orbits, where $\bar \lambda_z = \bar
J_z \times (\bar r / \bar \sigma)$. Here, $\bar J_z$ is the average
angular momentum along the $z$-direction and $\bar \sigma$ is the
average second moment of the orbit. NGC 1271 shows several dynamical
components, as can be seen in Figure \ref{fig:massdens}, including a
clear non-rotating bulge (with $-0.2 < \bar \lambda_z < 0.2$), a
highly co-rotating component (with $\bar \lambda_z \sim$0.5), and a
maximally co-rotating component (with $0.8 < \bar \lambda_z <
1.0$). The bulge component accounts for 12\% of the mass, whereas the
rotating components total 75\% of the mass within the radial extent of
the kinematic measurements. Qualitatively these components agree with
the classification of NGC 1271 as a fast-rotating S0 galaxy with a
classical bulge.

\subsection{Black Hole -- Host Galaxy Relations}
\label{subsec:bhrels}

Although the dynamical decomposition from the best-fit stellar
dynamical model presented above can be used to place NGC 1271 on the
$M_\mathrm{BH}$ -- bulge relationships, we follow the more common
approach of carrying out a photometric decomposition to determine the
galaxy's bulge luminosity and bulge effective radius. Using Galfit, we
find a single S\'{e}rsic component fit is an insufficient description
of the galaxy, with the percent difference between the model and data
reaching as high as 60\%. The fit is significantly improved with the
addition of one or two other S\'{e}rsic components, and in the later
case the percent difference between the model and data is under
15\%. In Table \ref{tab:galfitresults}, we present the best-fit
parameters of Galfit models with one, two, and three S\'{e}rsic
components, as well as the F160W luminosity for each component. The
three-component model provides the best match to the \emph{HST} image,
but it is difficult to unambiguously identify a ``bulge'' component
because the components all have rather low S\'{e}rsic
indices. Therefore, we conservatively assume that the innermost
component of the three-component model provides a lower limit on the
bulge luminosity and effective radius, while the innermost component
of the two-component model gives an upper limit. This yields a
$K$-band bulge luminosity of $(1.9 - 7.2)\times10^{10}\ L_\odot$ when
correcting for galactic extinction using the
\cite{Schlafly_Finkbeiner_2011} WFC3 F160W value of $0.085$, assuming
an $H-K$ color of $0.2$ \citep{Vazdekis_1996}, and a $K$-band solar
absolute magnitude of $3.29$. Alternatively, the bulge has a mass of
$(2.3-8.7)\times10^{10}\ M_\odot$ when applying the best-fit
mass-to-light ratio from our dynamical models to the luminosities in
Table \ref{tab:galfitresults}. The corresponding bulge effective
radius ranges between 0\farcs6 and 5\farcs2 (or $0.2 - 2.0$ kpc).

\begin{deluxetable}{ccccccccccc}
\tabletypesize{\scriptsize}
\tablewidth{0pt}
\tablecaption{Galfit Models \label{tab:galfitresults}}
\tablehead{
\colhead{Component} &
\colhead{} &
\colhead{$m_{H}$} &
\colhead{} &
\colhead{$L_{H}$ ($L_\odot$)} &
\colhead{} &
\colhead{$R_e$ (\arcsec)} &
\colhead{} &
\colhead{$n$} &
\colhead{} &
\colhead{$q^\prime$} \\
\colhead{(1)} &
\colhead{} &
\colhead{(2)} &
\colhead{} &
\colhead{(3)} &
\colhead{} &
\colhead{(4)} &
\colhead{} &
\colhead{(5)} &
\colhead{} &
\colhead{(6)}
}

\startdata

1   &&   10.72  &&   7.7$\times10^{10}$  &&   5.56  &&   4.78  &&  0.41 \\[0.5ex]

\hline
\\[-1.5ex]

1   &&   10.95  &&   6.2$\times10^{10}$  &&   5.23  &&   6.54  &&  0.54 \\
2   &&   12.47  &&   1.5$\times10^{10}$  &&   6.63  &&   0.93  &&  0.20 \\[0.5ex]

\hline
\\[-1.5ex]

1   &&   12.39  &&   1.6$\times10^{10}$  &&   0.61  &&   2.12  &&  0.68 \\
2   &&   11.80  &&   2.8$\times10^{10}$  &&   5.24  &&   1.17  &&  0.27 \\
3   &&   11.90  &&   2.6$\times10^{10}$  &&   10.46  &&   1.30  &&  0.62

\enddata

\tablecomments{Column (1) shows the S\'{e}rsic component number,
  column (2) provides the F160W apparent magnitude in the Vega system,
  not yet corrected for galactic extinction, column (3) gives the
  luminosity after a correction of $0.085$ for galactic extinction and
  assuming an absolute solar magnitude of $3.33$, column (4) is the
  effective radius, column (5) gives the S\'{e}rsic index, and column
  (6) lists the projected axis ratio.}

\end{deluxetable}

In order to determine the bulge stellar velocity dispersion for NGC
1271, we use the best-fit stellar dynamical model from Section
\ref{sec:results} and predict the luminosity-weighted second moment
within a circular aperture whose radius equals the galaxy's bulge
effective radius, following the approach used by
\cite{vandenBosch_2012}. Due to the uncertainty in the bulge effective
radius for NGC 1271, we measure the effective stellar velocity
dispersion ($\sigma_{e, \mathrm{bul}}$) for three different bulge
effective radii corresponding to the largest $R_e$ estimate from the
Galfit decomposition, the smallest estimate of $R_e$, and the average
of the two. Additionally, some previous black hole studies have chosen
to exclude data within $r_\mathrm{sphere}$ when determining
$\sigma_{e, \mathrm{bul}}$ because the stellar kinematics are under
the direct influence of the black hole in this region (e.g.,
\citealt{Gebhardt_2011, McConnell_Ma_2013}). When excluding the region
within $r_\mathrm{sphere}$, we find effective stellar velocity
dispersions of $\sigma_{e, \mathrm{bul}} = 272$ km s$^{-1}$,
$\sigma_{e, \mathrm{bul}} = 276$ km s$^{-1}$, and $\sigma_{e,
  \mathrm{bul}} = 349$ km s$^{-1}$, whereas when the region within
$r_\mathrm{sphere}$ is included we measure $\sigma_{e, \mathrm{bul}} =
285$ km s$^{-1}$, $\sigma_{e, \mathrm{bul}} = 294$ km s$^{-1}$, and
$\sigma_{e, \mathrm{bul}} = 358$ km s$^{-1}$, for bulge effective
radii of 5\farcs2, 2\farcs9, and 0\farcs6, respectively. As a
comparison, the HET Massive Galaxy Survey reports a central velocity
dispersion of 317 km s$^{-1}$ \citep{vandenBosch_2015}, which is the
observed stellar velocity dispersion within a 3\farcs5 aperture from
the major axis long-slit data.

\begin{figure*}
\begin{center}
\epsscale{0.95}
\plotone{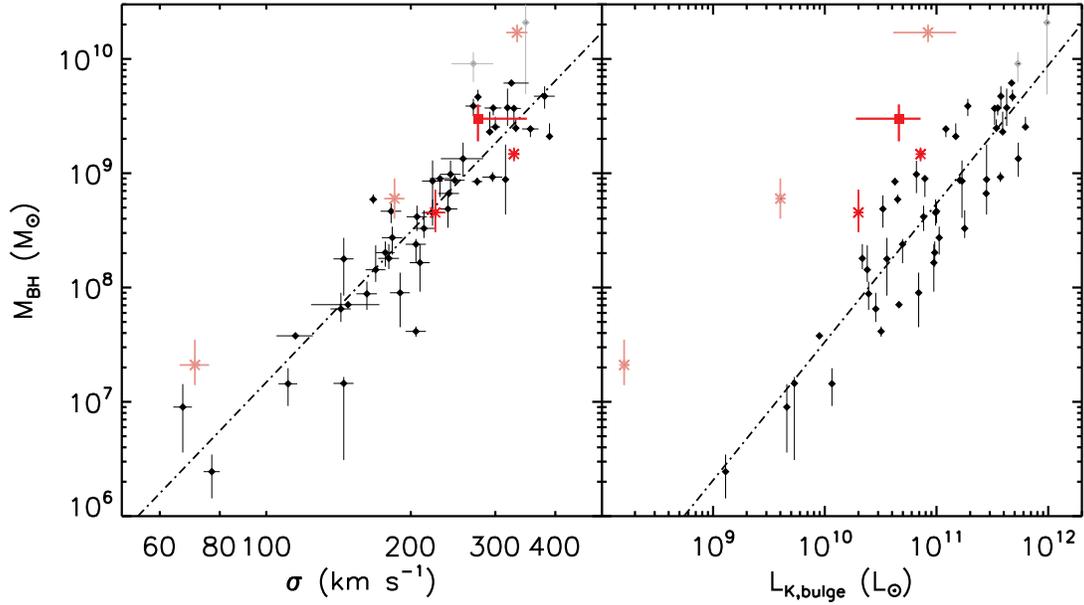}
\caption{NGC 1271 (red filled square) is shown on the black hole --
  host galaxy relations. The black hole/galaxy measurements (black and
  gray filled circles) and the fitted relations (dot dashed lines) are
  taken from \cite{Kormendy_Ho_2013}. The compact galaxies that have
  existing dynamical $M_\mathrm{BH}$ measurements are denoted with the
  red asterisks. These galaxies are generally consistent with
  $M_\mathrm{BH}$ -- $\sigma_\star$ but are positive outliers from
  $M_\mathrm{BH}$ -- $L_\mathrm{bul}$. \cite{Kormendy_Ho_2013} did not
  include the measurements shown in gray and light red when fitting
  the black hole scaling relations. \label{fig:bhrels}}
\end{center}
\end{figure*}

Figure \ref{fig:bhrels} shows the location of NGC 1271 on the most
recent versions of the $M_\mathrm{BH}-\sigma_\star$ and
$M_\mathrm{BH}-L_\mathrm{bul}$ relations by
\cite{Kormendy_Ho_2013}. For the purposes of placing NGC 1271 on the
$M_\mathrm{BH}$ -- $\sigma_\star$ correlation, we use $\sigma_\star =
276$ km s$^{-1}$ with uncertainties that include the $\sigma_{e,
  \mathrm{bul}}$ measurements made for bulge effective radii of
5\farcs2 and 0\farcs6 when excluding data within
$r_\mathrm{sphere}$. When placing NGC 1271 on the $M_\mathrm{BH}$ --
$L_\mathrm{bul}$ relation, we set the faint end of the bulge
luminosity error bar assuming the luminosity of the innermost
component of the three-component S\'{e}rsic fit to the \emph{HST}
image and the high end of the error bar assuming the luminosity of the
innermost component of the two-component S\'{e}rsic fit. We adopt a
$K$-band bulge luminosity of $4.6 \times 10^{10}\ L_\odot$, which is
the midpoint of the range of possible bulge luminosities.

\begin{figure*}
\begin{center}
\epsscale{0.8}
\plotone{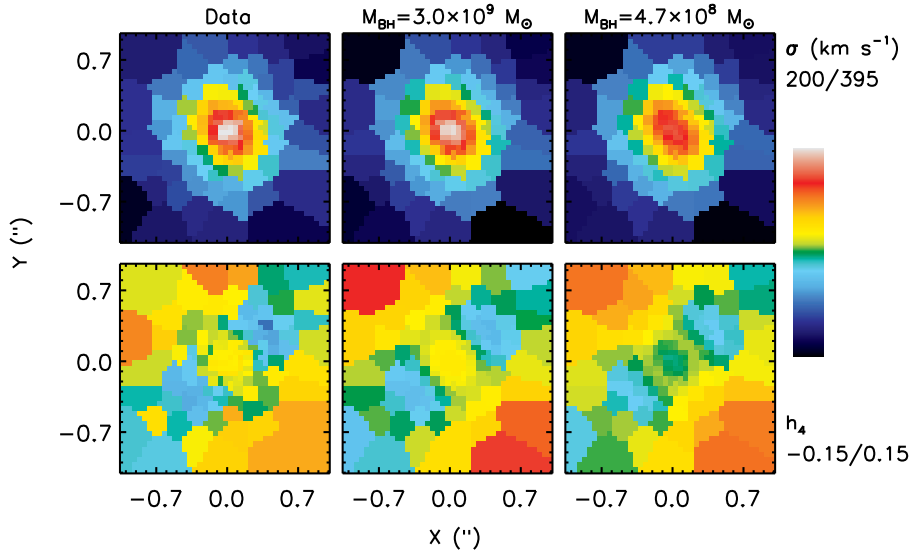}
\caption{The observed velocity dispersion (top) and $h_4$ (bottom)
  measured from the NIFS data (left) is compared to predictions from
  the best-fit model with $M_\mathrm{BH} = 3.0\times10^9\ M_\odot$
  (middle) and a model with a $4.7\times10^8\ M_\odot$ black hole
  (right), which is the mass predicted from $M_\mathrm{BH}$ --
  $L_\mathrm{bul}$ when conservatively adopting the total galaxy
  luminosity. When generating the $\sigma$ and $h_4$ predictions for a
  $4.7 \times 10^8\ M_\odot$ black hole, we sample over $\Upsilon_H$
  and the top five NFW dark halos from the model grid in Section
  \ref{sec:results}, such that the combination of parameters produces
  a model with the lowest $\chi^2$ for a black hole mass of $4.7
  \times 10^8\ M_\odot$. The data and model maps are plotted on the
  same scale, with the ranges given by the color bar to the right and
  the minimum and maximum values printed at the side of the maps. The
  best-fit model is able to reproduce the sharp rise in the velocity
  dispersion and slight peak in $h_4$ at the center, while the smaller
  black hole expected from the $M_\mathrm{BH}$ -- $L_\mathrm{bul}$
  correlation is unable to match either of these
  features. \label{fig:comparemodels}}
\end{center}
\end{figure*}

We find that NGC 1271 consistent with the $M_\mathrm{BH}$ --
$\sigma_\star$ relation, but is an order of magnitude above the black
hole mass prediction from the $M_\mathrm{BH}$ -- $L_\mathrm{bul}$
correlation. In order to demonstrate that the black hole in NGC 1271
must be larger than that expected from $M_\mathrm{BH} -
L_\mathrm{bul}$, in Figure \ref{fig:comparemodels} we present the NIFS
observations of the velocity dispersion and $h_4$ along with the
predictions from the best-fitting model with a black hole mass of $3.0
\times 10^9\ M_\odot$ and a $4.7 \times 10^8\ M_\odot$ black hole. The
$4.7 \times 10^8\ M_\odot$ black hole is expected from $M_\mathrm{BH}$
-- $L_\mathrm{bul}$ when conservatively using the galaxy's total
$K$-band luminosity of $8.9 \times 10^{10}\ L_\odot$, which is derived
from the single S\'{e}rsic fit to the \emph{HST} image, after
correcting for galactic extinction and assuming a $H-K=0.2$. Clear
differences between the kinematic predictions and the observations can
be seen by eye. The best-fit model with $M_\mathrm{BH} = 3.0 \times
10^9\ M_\odot$ is able to nicely reproduce the sharp rise in the
velocity dispersion and the slight peak in $h_4$ at the nucleus, while
the less massive black hole predicted from $M_\mathrm{BH}$ --
$L_\mathrm{bul}$ fails to do so.

NGC 1271 has an apparent ellipticity of $\epsilon=0.6$ and a specific
stellar angular momentum of $\lambda_R \equiv \langle R | V | \rangle
/ \langle R \sqrt{V^2 + \sigma^2} \rangle = 0.5$ within one effective
radius based on the PPAK data. Here, $R$, $V$, and $\sigma$ are the
radius, velocity, and velocity dispersion and the brackets denote a
luminosity weighted average \citep{Emsellem_2007}. Using the dividing
line between slow and fast rotators from the ATLAS$^{3\mathrm{D}}$
survey, such that fast rotators have $\lambda_R \geq 0.31 \times
\sqrt{\epsilon}$ \citep{Emsellem_2011}, NGC 1271 falls well within in
this fast rotator regime.

NGC 1271 appears similar to the other compact galaxies NGC 1277, NGC
1332, NGC 4342, NGC 4486B, and M60-UCD1. All six of these early-type
galaxies have small sizes, are rotating, show large stellar velocity
dispersions for their luminosities, and have black holes that are too
massive for their host galaxy's bulge luminosity. The black holes,
however, are consistent with $M_\mathrm{BH}$ -- $\sigma_\star$ given
the intrinsic scatter of the relation. In the case of M60-UCD1,
\cite{Seth_2014} suggest that the ultracompact dwarf galaxy (UCD) was
once the nucleus of a larger galaxy that has since been tidally
stripped by the giant elliptical M60, whose center lies at a projected
distance of just 6.6 kpc away from the UCD. While tidal stripping is a
natural explanation for the presence of an over-massive black hole, in
the case of NGC 1271, we do not see signs of active stripping in the
\emph{HST} image. The isophotes appear extremely regular, and no
massive galaxies immediately neighbor NGC 1271 like in the case of
M60-UCD1. NGC 1271 is $\sim$270 kpc in projection from the BCG of
Perseus. Further evidence could come from counting the number of
globular clusters, as the galaxy would be stripped of its globular
clusters first. While NGC 1271 appears not to have been stripped with
our current data, we cannot rule out an event in the distant past.

Interestingly, the behavior of the compact, high-dispersion galaxies
being consistent with $M_\mathrm{BH}$ -- $\sigma_\star$ but being
large positive outliers on $M_\mathrm{BH}$ -- $L_\mathrm{bul}$ could
be in conflict with recent observations of BCGs, which instead may
hint that black hole mass becomes independent of $\sigma_\star$ at
high black hole mass while the $M_\mathrm{BH}$ -- $L_\mathrm{bul}$
correlation remains unchanged at large luminosities
\citep{McConnell_Ma_2013, Kormendy_Ho_2013}. Clearly, more compact,
high-dispersion galaxies and BCGs/giant ellipticals need to be
examined. There could be systematic differences in the scaling
relations between the two types of galaxies, thereby imply different
mechanisms for black hole growth. Since the compact, high-dispersion
galaxies like NGC 1271 look similar to the quiescent $z\sim2$ red
nuggets, they could be relics that somehow avoided the same fate that
ultimately produced the giant ellipticals observed today. Perhaps the
compact, high-dispersion galaxies are left over from an era when the
local black hole scaling relations did not apply and galaxies instead
contained over-massive black holes.

\section{Conclusion}
\label{sec:conclusion}

To summarize, we obtained AO-assisted Gemini NIFS observations of NGC
1271 to map out the stellar kinematics on scales comparable to the
black hole sphere of influence, and large-scale IFU data with PPAK,
which are useful for constraining the galaxy's stellar mass-to-light
ratio and orbital distribution. Using an \emph{HST} WFC3 $H$-band
image along with the spectral information, we constructed orbit-based
stellar dynamical models. We measure $M_\mathrm{BH} =
(3.0^{+1.0}_{-1.1})\times10^9\ M_\odot$ and $\Upsilon_H =
1.40^{+0.13}_{-0.11}\ \Upsilon_\odot$. The quoted errors combine the
1$\sigma$ model fitting uncertainties with some common sources of
systematic uncertainty that affect stellar dynamical models. The black
hole in NGC 1271 is at the upper end of the black hole mass
distribution ($M_\mathrm{BH} > 1\times10^9\ M_\odot$). Yet, this
compact, rapidly rotating galaxy, with a high stellar velocity
dispersion for its luminosity is very different from the giant
elliptical galaxies and BCGs that are expected to harbor the most
massive black holes in the Universe. Such host galaxy environments
have yet to be widely explored on the $M_\mathrm{BH}$ -- host galaxy
relations. With our mass measurement, we find that the black hole is
too large for the galaxy's $K$-band bulge luminosity of
$(4.6^{+2.6}_{-2.7})\times10^{10}\ L_\odot$, falling an order of
magnitude above the expectation from the $M_\mathrm{BH}$ --
$L_\mathrm{bul}$ correlation, but the black hole mass is consistent
with expectations from the $M_\mathrm{BH}$ -- $\sigma_\star$
relationship assuming $\sigma_\star = 276^{+73}_{-4}$ km
s$^{-1}$. This behavior has also been observed in the few other
compact galaxies that have dynamical black hole mass measurements to
date. Carrying out more black hole mass measurements in similar
galaxies using high spatial resolution observations from \emph{HST}
and AO is necessary in order to determine if there are systematic
differences in the black hole scaling relations between the large
ellipticals/BCGs and these compact, high-dispersion galaxies. The
compact, high-dispersion galaxies could be remnants of the $z\sim2$
red nugets that for some reason did not evolve into the largest
ellipticals observed today, and instead reflect a time when black
holes were too large for their bulges. More broadly, additional black
hole mass measurements are needed in order to enlarge and better fill
in undersampled regions of galaxy parameter space. Targeting such a
large and carefully selected sample with high spatial resolution
facilities is a natural step toward gaining a more complete census of
local black holes and a better understanding of the role that black
holes play in galaxy evolution.

\acknowledgements

J.~L.~W. has been supported by an NSF Astronomy and Astrophysics
Postdoctoral Fellowship under Award No. 1102845. Based on observations
obtained at the Gemini Observatory, which is operated by the
Association of Universities for Research in Astronomy, Inc., under a
cooperative agreement with the NSF on behalf of the Gemini
partnership: the National Science Foundation (United States), the
National Research Council (Canada), CONICYT (Chile), the Australian
Research Council (Australia), Minist\'{e}rio da Ci\^{e}ncia,
Tecnologia e Inova\c{c}\~{a}o (Brazil) and Ministerio de Ciencia,
Tecnolog\'{i}a e Innovaci\'{o}n Productiva (Argentina), under program
GN-2012B-Q-51. Also based on observations made with the NASA/ESA
Hubble Space Telescope, obtained at the Space Telescope Science
Institute, which is operated by the Association of Universities for
Research in Astronomy, Inc., under NASA contract NAS 5-26555. These
observations are associated with program \#13050. This work is further
based on observations collected at the Centro Astron\'{o}mico Hispano
Alem\'{a}n (CAHA) at Calar Alto, operated jointly by the Max-Planck
Institut f\"{u}r Astronomie and the Instituto de Astrof\'{i}sica de
Andaluc\'{i}a (CSIC). The authors acknowledge the Texas Advanced
Computing Center (TACC; http://www.tacc.utexas.edu) at The University
of Texas at Austin for providing HPC resources that have contributed
to the research results reported within this paper. The authors also
made use of the grant-funded cyberinfrastructure at Indiana
University. This material is based upon work supported by the National
Science Foundation under Grant No. CNS-0723054, and in part by Lilly
Endowment, Inc., through its support for the Indiana University
Pervasive Technology Institute, and in part by the Indiana METACyt
Initiative. The Indiana METACyt Initiative at Indiana University is
also supported in part by Lilly Endowment, Inc. This research has made
use of the NASA/IPAC Extragalactic Database which is operated by the
Jet Propulsion Laboratory, California Institute of Technology, under
contract with NASA.

\end{document}